\documentclass[twocolumn,showkeys,showpacs,preprintnumbers,prd,superscriptaddress,nofootinbib]{revtex4-2}
\usepackage{graphicx}
\usepackage{array}
\usepackage{epsf}
\usepackage{bm}
\usepackage{amsmath}
\usepackage{amsfonts}
\usepackage{amssymb}
\usepackage{epstopdf}
\usepackage{natbib}
\usepackage{hyperref}
\usepackage{color}
\usepackage{verbatim}
\usepackage{multirow}
\usepackage{bm}
\usepackage{hyperref}
\usepackage{float}
\usepackage{tabularx}
\usepackage[normalem]{ulem}

\DeclareUnicodeCharacter{2212}{-}
\definecolor{darkblue}{rgb}{0.0, 0.0, 0.55}
\definecolor{darkred}{rgb}{0.55, 0.0, 0.0}

\usepackage{hyperref}
\hypersetup{
    colorlinks=true, 
    linkcolor=darkblue,
    citecolor=darkblue,
    urlcolor=darkblue}
    
\makeatletter\let\expandableinput\@@input\makeatother

\newcolumntype{Y}{>{\centering\arraybackslash}X}
\renewcommand{\arraystretch}{1.75}

\newcommand{\Tstrut}{\rule{0pt}{3.2ex}}        
\newcommand{\Bstrut}{\rule[-3.2ex]{0pt}{0pt}} 

\begin{document}

\title{Joint Constraints on Neutrinos and Dynamical Dark Energy in Minimally Modified Gravity}


\author{Artur Ladeira}
\email{artur.ladeira@ufrgs.br}
\affiliation{Instituto de F\'{i}sica, Universidade Federal do Rio Grande do Sul, 91501-970 Porto Alegre RS, Brazil}

\author{Rafael C. Nunes}
\email{rafadcnunes@gmail.com}
\affiliation{Instituto de F\'{i}sica, Universidade Federal do Rio Grande do Sul, 91501-970 Porto Alegre RS, Brazil}
\affiliation{Divisão de Astrofísica, Instituto Nacional de Pesquisas Espaciais, Avenida dos Astronautas 1758, São José dos Campos, 12227-010, São Paulo, Brazil}

\author{Supriya Pan}
\email{supriya.maths@presiuniv.ac.in}
\affiliation{Department of Mathematics, Presidency University, 86/1 College Street, Kolkata 700073, India}
\affiliation{Institute of Systems Science, Durban University of Technology, Durban 4000, Republic of South Africa}

\author{Weiqiang Yang}
\email{d11102004@163.com}
\affiliation{Department of Physics, Liaoning Normal University, Dalian, 116029, P. R.  China}

\begin{abstract}
The \(w_{\dagger}\)VCDM framework provides a theoretically well-controlled extension of \(\Lambda\)CDM within the class of minimally modified gravity theories, allowing for flexible cosmological background evolution and linear perturbation dynamics while remaining free of pathological instabilities. In this work, we have shown that this scenario remains robust when confronted with current cosmological observations, even in the presence of an extended neutrino sector. Combining \textit{Planck} CMB data with DESI DR2 BAO and PantheonPlus, we obtain stringent constraints on neutrino physics, including \(\sum m_\nu < 0.12~\mathrm{eV}\) (95\% CL) and \(N_{\rm eff} = 3.114^{+0.139}_{-0.128}\), fully consistent with Standard Model expectations within $1\sigma$. Crucially, the data exhibit a statistically significant preference for a late-time dark-energy transition, characterized by a robust quintessence--phantom crossing that remains stable across all dataset combinations and neutrino-sector extensions, including the presence of a sterile neutrino. The combined effects of modified late-time expansion and additional relativistic degrees of freedom systematically raise the inferred Hubble constant, substantially alleviating the \(H_0\) tension without invoking early dark energy or introducing theoretical instabilities. Overall, the \(w_{\dagger}\)VCDM scenario emerges as a compelling phenomenological framework that simultaneously accommodates current constraints on neutrino physics, provides an excellent fit to recent BAO and supernovae data, and offers a viable pathway toward resolving persistent tensions in the standard cosmological model.

\end{abstract}

\maketitle
\section{Introduction}

The standard cosmological model, $\Lambda$ Cold Dark Matter ($\Lambda$CDM), provides a minimal and highly successful description of the present Universe. In this framework, dark energy (DE) is modeled as a cosmological constant with a fixed equation-of-state (EoS) $w=-1$, while dark matter (DM) is pressureless. Despite its remarkable agreement with a wide range of cosmological observations, $\Lambda$CDM faces several persistent challenges that may point to a new physics beyond the concordance model (see \cite{DiValentino:2021izs,Perivolaropoulos:2021jda,Kamionkowski:2022pkx} for a review).

Among these, the Hubble constant ($H_0$) tension remains one of the most significant challenges to modern cosmology. The discrepancy between early- and late-Universe determinations of $H_0$ has reached a statistical significance of $\sim 5\sigma$~\cite{Riess:2021jrx}, and recent analyses by the H0DN collaboration suggest that it may increase to $7.1\sigma$ \cite{H0DN:2025lyy}. If confirmed, such a discrepancy would be difficult to accommodate within the standard $\Lambda$CDM framework. Independent tensions have also emerged from recent measurements of baryon acoustic oscillations (BAO) by the Dark Energy Spectroscopic Instrument (DESI)~\cite{DESI:2024mwx,DESI:2025zgx}. In its first two data releases (DR1 and DR2), DESI team reports indications that the DE component may exhibit time evolution rather than a strictly constant energy density ~\cite{DESI:2024mwx,DESI:2025zgx}. These findings provide an additional and complementary challenge to the $\Lambda$CDM paradigm.

In particular, the DESI DR2 BAO data indicate a preference for dynamical DE at the $2.8\sigma$--$4.2\sigma$ level,
with the reconstructed EoS exhibiting a transition from a phantom-like regime at earlier times to a quintessence-like behavior at late times. This evidence represents a nontrivial challenge to the concordance $\Lambda$CDM model. Moreover, several independent analyses have confirmed significant deviations from the \(\Lambda\)CDM framework and proposed new cosmological tests to further probe it in light of DESI-DR2 BAO samples~\cite{DESI:2025zgx,DESI:2025zpo,Montani:2025qnk,Capozziello:2025qmh,Ozulker:2025ehg,deCruzPerez:2025dni,Yao:2025wlx,Sohail:2025mma,Smith:2025uaq,Lee:2025axp,Efstratiou:2025iqi,Smith:2025icl,Liu:2025bss,Li:2025muv,RoyChoudhury:2025iis,Chen:2025ywv,Fazzari:2025lzd,Gomez-Valent:2025mfl,Chaudhary:2025vzy,SanchezLopez:2025uzw,Artola:2025srt,Artola:2025zzb,Wolf:2025acj,Adam:2025kve,Wu:2025vfs,Feleppa:2025clx,Mishra:2025goj,Hussain:2025vbo,Gialamas:2025pwv,Cline:2025sbt,Mukherjee:2025ytj,Bayat:2025xfr,Hussain:2025nqy,Cheng:2025lod,Toomey:2025yuy,Paliathanasis:2025kmg,Luciano:2025hjn,Li:2025vuh,Cheng:2025yue} or alternative datasets~\cite{Silva:2025twg,Sabogal:2025jbo,Chudaykin:2025lww,Reeves:2025xau,Ishak:2025cay}. 

On the other hand, cosmological observations provide a powerful and complementary probe of the absolute neutrino mass scale (see, e.g., Refs.~\cite{Lesgourgues:2006nd,Lesgourgues:2014zoa,Vagnozzi:2019utt,DiValentino:2024xsv,Escudero:2024uea} for reviews), offering sensitivity beyond that accessible to laboratory experiments. Massive neutrinos leave characteristic imprints on cosmological observables by affecting the expansion history, delaying the growth of cosmic structures, and suppressing power on small scales due to free streaming. The inference of neutrino masses from cosmological data, however, is inherently model dependent. Degeneracies with other cosmological parameters, particularly those associated with DE—can significantly alter the derived bounds when extensions of the $\Lambda$CDM model are considered \cite{Yang:2017amu,Yang:2019uog,Yang:2020ope,Liu:2020vgn,DiValentino:2021rjj,diValentino:2022njd,Pan:2023frx,Du:2024pai,Bertolez-Martinez:2024wez,Wang:2024hen,Gariazzo:2024sil,Shao:2025ohz,Ghedini:2025epp,Nair:2025uyn,Chudaykin:2025lww,Barua:2025adv,Du:2025xes,Wang:2025ker,Feng:2025mlo,Zhou:2025nkb}. Recent analyses using state-of-the-art datasets, including BAO from DESI DR2, have provided stringent constraints on neutrino properties within both minimal and some extended cosmological frameworks (see, e.g., Refs.~\cite{Jiang:2024viw,Gariazzo:2022ahe,Jiang:2024viw,RoyChoudhury:2024wri,Giare:2025ath,Wang:2025ker,RoyChoudhury:2025dhe,Elbers:2025vlz,Du:2025xes}).

In light of the emerging indications for dynamical DE from recent joint cosmological analyses, it becomes essential to reassess cosmological constraints on neutrino masses beyond the $\Lambda$CDM paradigm, explicitly accounting for potential degeneracies between neutrino properties and nonstandard DE dynamics. Such correlations may play a crucial role in shaping the inferred bounds on neutrino parameters and in interpreting apparent departures from the concordance model.

In this work, we focus on the late-time dynamical behavior of DE, with particular emphasis on recent evidence for a possible phantom crossing within the VCDM\footnote{The term VCDM is used to denote the scenario in which $\Lambda$ of $\Lambda$CDM is replaced by a time-dependent potential $V$. } framework ~\cite{Scherer:2025esj}. This model belongs to the class of minimally modified gravity theories and has been shown to provide an improved fit to current cosmological datasets. By combining the most recent cosmological observations—including DESI-DR2 BAO measurements, Type Ia supernova samples from PantheonPlus, DES-Y5, and Union3, together with Planck CMB data—robust statistical support is found for a transition between quintessence and phantom dark-energy behaviors across multiple dataset combinations. The most pronounced indication arises from the Planck+DESI+DES-Y5 dataset, which shows a preference above the $\sim5\sigma$ level for a transition occurring at a redshift $z_\dagger \approx 0.49$. Other analyses presented in the same work lead to consistent conclusions regarding the value of $z_\dagger$. Although the inferred equation of state evolves from a phantom-like regime at higher redshifts to a quintessence-like behavior at later times, this dynamical evolution does not resolve the $H_0$ tension.
Recently, the minimally modified gravity model VCDM has been extended and tested against well-known parameterizations in the literature~\cite{Arora:2025msq}.

Motivated by the central role of neutrinos in precision cosmology, we extend previous analyses of this scenario by incorporating massive neutrinos and exploring their degeneracies with the parameters governing DE dynamics. Our study aims to assess the robustness of the VCDM phenomenology when neutrino masses are allowed to vary, and to clarify the interplay between neutrino properties and late-time cosmic acceleration in models that exhibit significant evidence beyond the $\Lambda$CDM framework.

This article is organized as follows. In Sec.~\ref{model}, we briefly review the theoretical framework of the model previously introduced in Ref.~\cite{Scherer:2025esj}. In Sec.~\ref{data}, we describe the datasets and the methodology employed in our analysis. Our main results and their discussions are presented in Sec.~\ref{results}. Finally, we summarize our findings and outline future prospects in Sec.~\ref{final}.

\section{The $w_{\dagger}$VCDM parametrization}
\label{model}

Minimally modified gravity provides a powerful theoretical framework for extending General Relativity without introducing additional physical degrees of freedom~\cite{Aoki:2018brq,DeFelice:2020eju,DeFelice:2015hla}. In this work, we adopt the VCDM framework~\cite{DeFelice:2020eju}. Within this approach, the cosmological constant $\Lambda$ appearing in the standard $\Lambda$CDM scenario is promoted to a generalized function $V(\phi)$, where $\phi$ is an auxiliary field that does not represent a propagating dynamical degree of freedom. The designation VCDM reflects the introduction of this variable potential $V(\phi)$ as the driver of late-time cosmic acceleration.

By construction, VCDM belongs to the class of minimal theories of gravity, meaning that it preserves the number of local physical degrees of freedom present in General Relativity. As a result, the theory does not introduce additional scalar or vector gravitational modes beyond the usual tensor polarizations of gravitational waves. This feature plays a central role in ensuring the theoretical consistency of the model and in avoiding common pathologies encountered in extended gravity scenarios. In particular, the absence of extra propagating modes guarantees that ghost-like instabilities are avoided at both the background and perturbative levels.

An additional advantage of the VCDM framework is that it allows for non-trivial background cosmological dynamics without jeopardizing stability. Many alternative dark energy or modified gravity models suffer from instabilities when attempting to produce departures from $\Lambda$CDM at the background level. In contrast, the minimal construction of VCDM enables a wider range of phenomenological behaviors for the dark energy sector, while maintaining a well-controlled and stable perturbative structure. This makes VCDM a particularly suitable theoretical setting for exploring departures from a constant equation of state and for testing extended dark energy scenarios against current cosmological data. Within this framework, the equations governing the dynamical evolution of the model are summarized in Sec.~II of Ref.~\cite{Arora:2025msq}. The original gravitational formulation of the theory was introduced in~\cite{DeFelice:2020eju}.

To set the stage for our analysis, we revisit the theoretical construction introduced in~\cite{Scherer:2025esj}. In that work, a phenomenological yet powerful diagnostic was developed to probe the dynamics of DE. The method not only places constraints on the DE EoS but also tests whether current data exhibit a statistical preference for a transition between quintessence-like ($w > -1$) and phantom-like ($w < -1$) regimes—or the reverse. This approach provides a minimally parametric way to investigate possible departures from $\Lambda$CDM while ensuring that the transition is triggered solely by observational evidence rather than theoretical bias.

A key advantage of this setup is that the corresponding dynamical behavior can be embedded, in a theoretically robust manner, within the VCDM framework~\cite{DeFelice:2020eju}. VCDM constitutes a broad class of DE models capable of reproducing a rich variety of time-dependent EoS evolutions (see also ~\cite{DeFelice:2020cpt,Arora:2025msq}). Importantly, it achieves this without introducing any additional propagating degrees of freedom in either the gravitational or matter sectors. As a result, the model is free from the typical instabilities—such as ghosts or gradient instabilities—that frequently arise in non-canonical or modified-gravity constructions. At the background level, the theory is internally consistent, and the linear perturbation sector remains equally well behaved.

The possibility of embedding a $w$-switch into VCDM further strengthens the physical viability of this phenomenological test. Within VCDM, the EoS evolution can be interpreted as arising from an effective modification of the gravitational sector that modifies the background expansion while leaving the number of propagating fields unchanged. This makes the framework particularly attractive for exploring non-trivial DE dynamics without jeopardizing fundamental consistency. Theoretical investigations of VCDM have already been carried out across a wide range of contexts, including compact-object physics and cosmological evolution~\cite{DeFelice:2021xps, DeFelice:2022riv, Jalali:2023wqh, DeFelice:2022uxv, DeFelice:2020cpt}. In addition, several dynamical DE scenarios have been successfully embedded in this framework~\cite{Akarsu:2024qsi, Akarsu:2024eoo, CosmoVerseNetwork:2025alb}, demonstrating its flexibility and general applicability. These previous studies provide the theoretical foundation upon which we build the observational analysis presented here.

Since the relevant theoretical developments are already well established in the literature, here we simply summarize the two key differences in the background and perturbation dynamics of this model relative to $\Lambda$CDM. First, regarding the background evolution, the transition in the EoS must be smooth. In general, the model assumes that the EoS undergoes a transition at a critical redshift, with a parameter $\Delta$ controlling the amplitude of the change. The transition is centered around a critical scale factor $a_{\dagger}$ (or, equivalently, a redshift $z_{\dagger}$) and represents a rapid yet continuous variation in the value of the DE EoS. To capture this behavior in a smooth and analytically convenient way, we adopt the following functional form for $w(N)$:
\begin{equation}
\label{EoS_smooth}
    w(N) = -1 + \Delta \tanh\!\left[\zeta (N_\dagger - N)\right],
\end{equation}
where the parameter $\zeta$ determines the sharpness of the transition and $N$ denotes the e-folding time, defined through $N = \ln(a/a_0)$, such that $N \leq 0$ at all epochs of interest. Larger values of $\zeta$ correspond to a steeper and more sudden transition, whereas smaller values generate a more gradual evolution of $w(a)$ across the critical epoch.

Throughout all analyses conducted in this work, the parameter $\zeta$ is fixed to $10^{1.5}$. The parameter $\zeta$ controls the smoothness of the transition. However, preliminary tests performed in the course of this and previous works indicate that varying $\zeta$ across several orders of magnitude produces no statistically significant changes in the posterior distributions of the main baseline parameters ($H_0$, $\Omega_m$, $S_8$, $\Delta$, $z_\dagger$, etc.). This lack of sensitivity arises because, for all currently available cosmological probes---in particular, combinations of CMB and BAO data---the dominant constraining power is associated with the location of the transition rather than with its sharpness. Consequently, all main results presented in this paper are effectively independent of the chosen value of $\zeta$, in agreement with both physical expectations and observational intuition.

We briefly review the expansion history in the presence of neutrinos. In what follows, all numerical and statistical analyses are performed using the Boltzmann solver \textsc{CLASS} code~\cite{Blas:2011rf}, which is employed throughout the main results of this paper. In the presence of both massless and massive neutrinos, the Hubble expansion rate is determined by the Friedmann equation,
\begin{equation}
H^{2}(a)=\frac{8\pi G}{3} \rho_{\rm tot}(a),
\end{equation}
where the total energy density is given by
\begin{equation}
\rho_{\rm tot}(a)=\rho_\gamma(a)+\rho_\nu(a)+\rho_{m}(a)+\rho_{\rm DE}(a).
\end{equation}
Here, $\rho_\gamma(a)$ and $\rho_\nu(a)$ denote the photon and neutrino energy densities, respectively, $\rho_{\rm DE}(a)$ corresponds to the energy density of the DE component, and $\rho_{m}(a)$ represents the total matter contribution, including both cold dark matter and baryons.

The neutrino sector requires special treatment because its energy density interpolates between a radiation-like regime at early times and a matter-like regime at late times. For relativistic neutrinos (\(T_\nu \gg m_\nu\)), the energy density is
\begin{equation}
\rho_{\nu}^{\rm rel}(a)
= N_{\rm eff}\,\frac{7}{8}\left(\frac{4}{11}\right)^{4/3}\rho_{\gamma 0}\,a^{-4},
\end{equation}
with \(N_{\rm eff}\) denoting the effective number of relativistic species.
For massive neutrinos, the exact energy density is
\begin{equation}
\rho_\nu(a)=\frac{T_{\nu 0}^{4}}{2\pi^{2}a^{4}}
\sum_{i}\int_{0}^{\infty}dq \, q^{2}
\frac{\sqrt{q^{2}+\left(a\,m_{\nu_i}/T_{\nu 0}\right)^{2}}}{e^{q}+1},
\end{equation}
which describes the transition from the ultrarelativistic regime (\(\rho_\nu\propto a^{-4}\)) to the non-relativistic regime (\(\rho_\nu\propto a^{-3}\)). 
The Friedmann equation can therefore be written in a practical form as
\begin{equation}
H^{2}(a)
=H_{0}^{2}\Big[
\Omega_{\gamma}\,a^{-4}
+\Omega_{\nu}(a)
+\Omega_{m}\,a^{-3}
+\Omega_{\rm DE}(a)
\Big],
\end{equation}
where \(\Omega_{\nu}(a)\) includes both the relativistic and massive neutrino contributions, and may also incorporate additional sterile-like species if present. Moreover, $\Omega_{\gamma}$, $\Omega_m$ and $\Omega_{\rm DE}$ respectively denotes the density parameter corresponding to radiation, matter and DE sector at the present moment. 

As emphasized in Ref.~\cite{Scherer:2025esj}, the dynamics of scalar linear perturbations in this framework closely resemble those of the standard $\Lambda$CDM model. The only modification appears in the momentum constraint equation, which captures the effects of the time-dependent DE sector and can be written as

\begin{equation}
\dot{\Phi}+aH\Psi =
\frac{3,[k^{2}-3a^{2}(\dot{H}/a)]\sum_{I}(\varrho_{I}+p_{I}),\theta_{I}}{k^{2},[2k^{2}/a^{2}+9\sum_{K}(\varrho_{K}+p_{K})]},
\label{eqn:dotphi}
\end{equation}
where $\Phi$ and $\Psi$ denote the two Bardeen potentials, $\theta_I$ is the scalar velocity perturbation of each fluid component, and the indices $I$ and $K$ run over all standard matter species.

The key difference relative to $\Lambda$CDM arises from the presence of the conformal-time derivative of the Hubble parameter, $\dot{H}/a$, which acquires an additional contribution due to the non-trivial dynamics of DE sector. Importantly, despite this modification, the theory remains free from physical pathologies: no ghost, gradient, or tachyonic instabilities arise, either at the background level or within linear perturbation theory. The model therefore provides a theoretically consistent and controlled framework in which to investigate non-standard DE dynamics.

When $\Delta = 0$, the $w_{\dagger}$VCDM model reduces exactly to the standard $\Lambda$CDM scenario, both in the background evolution and in the linear perturbation sector. In this limit, the expansion history and the growth of cosmic structures are identical to those predicted by $\Lambda$CDM, ensuring a smooth and continuous connection with the standard cosmological model. The only distinction between $\Lambda$CDM and this class of VCDM models is the additional parameter $\Delta$, which quantifies deviations from the standard cosmological framework. Non-zero values of $\Delta$ introduce modifications simultaneously at the background level and in the linear perturbation sector. Therefore, $\Delta$ directly controls the departure from $\Lambda$CDM, and the model smoothly reduces to $\Lambda$CDM in the limit $\Delta = 0$.

With this setup established, we now turn to the observational analysis and investigate the constraints on the $w_{\dagger}$VCDM class of models in the following sections.

\section{Datasets and methodology}
\label{data}

This section presents the observational datasets and the statistical methodology employed to constrain the proposed cosmological scenarios in the presence of massive neutrinos. We begin by introducing the key datasets used throughout this analysis:

\begin{enumerate}

\item {\bf Cosmic Microwave Background (CMB):} Measurements of temperature and polarization anisotropy of the CMB power spectra from Planck 2018 release alongside their cross-spectra and  CMB lensing measurements have been considered~\cite{Planck:2018vyg} . In particular, we consider high-$\ell$ \texttt{Plik} likelihood for TT (covering the multipole range $30 \leq \ell \leq 2508$), TE, and EE ($30 \leq \ell \leq 1996$) with the low-$\ell$ TT-only ($2 \leq \ell \leq 29$) likelihood and the low-$\ell$ EE-only ($2 \leq \ell \leq 29$) \texttt{SimAll} likelihood~\cite{Planck:2019nip}. CMB lensing measurements are reconstructed from the temperature 4-point correlation function~\cite{Planck:2018lbu}. This dataset is collectively referred to as \textbf{Planck}.

\item \textbf{Baryon Acoustic Oscillations} (\textbf{DESI-DR2}):  
Baryon acoustic oscillation (BAO) measurements from the second data release of the Dark Energy Spectroscopic Instrument (DESI) have been considered. This dataset includes BAO signals extracted from galaxy and quasar samples~\cite{DESI:2025zgx}, and as well as from the Lyman-$\alpha$ forest tracers~\cite{DESI:2025zpo}. The measurements, summarized in Table~IV of Ref.~\cite{DESI:2025zgx}, span the effective redshift interval $0.295 \leq z \leq 2.330$, and are binned into nine redshift slices. The BAO observables are provided in terms of the transverse comoving distance $D_{\mathrm{M}}/r_d$, the Hubble distance $D_{\mathrm{H}}/r_d$, and the spherically averaged distance $D_{\mathrm{V}}/r_d$, all normalized to the sound horizon at the baryon drag epoch, $r_d$. Correlations among these quantities are fully taken into account through the corresponding covariance matrix, including the cross-correlation coefficients $r_{\rm V,M/H}$ and $r_{\rm M,H}$. Throughout this work, this dataset is denoted as \textbf{DR2}.

\item \textbf{Type Ia Supernovae} (\textbf{SNIa}):  
We consider two distinct samples of SNIa as follows: (i)  PantheonPlus sample~\cite{Brout:2022vxf} comprising 1701 light-curve measurements of 1550 distinct SNIa distributed in $0.01 \leq z \leq 2.26$ (this dataset is denoted as \textbf{PP}),
(ii)  Union~3.0 compilation~\cite{Rubin:2023jdq} containing 2087 SNIa in $0.001 < z < 2.26$, with 1363 objects in common with the PantheonPlus sample(labeled as \textbf{Union3}).\footnote{This dataset employs a Bayesian hierarchical framework to consistently account for systematic uncertainties and measurement errors.}


\end{enumerate}

In the main analysis carried out in Ref.~\cite{Scherer:2025esj}, the total mass of the three active neutrino species was fixed to the minimum value allowed by neutrino oscillation experiments, $\sum m_\nu = 0.06,\mathrm{eV}$, consistent with the normal mass ordering. In this fiducial setup, the neutrino sector is modeled by one massive and two effectively massless states, an approximation that is commonly adopted and sufficiently accurate for standard cosmological analyses.

In this work, we extend the baseline analysis by considering the following generalized scenarios:

\begin{itemize}
\item \textbf{$w_{\dagger}$VCDM + $\sum m_\nu$}:
In this scenario, the total mass of the three active neutrino species, $\sum m_\nu$, is treated as a free parameter, while the effective number of relativistic species is fixed to its standard value, $N_{\rm eff} = 3.042$. This setup allows us to assess the impact of varying the absolute neutrino mass scale within the $w_{\dagger}$VCDM framework, while preserving the standard thermal history.

\item \textbf{$w_{\dagger}$VCDM + $\sum m_\nu + N_{\rm eff}$}:
Here, both the total neutrino mass and the effective number of relativistic degrees of freedom are allowed to vary. This extension captures possible deviations from the standard neutrino sector, including scenarios with additional relativistic species or nonstandard thermal histories, and enables a systematic study of degeneracies between neutrino properties and DE dynamics.

\item \textbf{$w_{\dagger}$VCDM + sterile neutrino}:
Finally, we explore the phenomenological implications of introducing an additional sterile neutrino species. Although the presence of sterile neutrinos is increasingly disfavored by recent laboratory experiments and cosmological analyses, this scenario is included to test the robustness of the $w_{\dagger}$VCDM framework against nonminimal neutrino sectors and to quantify the impact of such extensions on the inferred cosmological constraints.
\end{itemize}

In order to constrain the parameter space of the proposed scenarios, we modified the publicly available cosmological Boltzmann code \texttt{CLASS}~\cite{Blas:2011rf} to incorporate the background evolution and linear perturbation regime (scalar modes) of the extended $w_{\dagger}$VCDM model.
We have used the publicly available sampler \texttt{Cobaya}~\cite{Torrado:2020dgo} to extract the constraints. The convergence of the chains have been ensured through the Gelman-Rubin criterion ~\cite{Gelman_1992} with $R-1 < 10^{-2}$. In Table~\ref{tab:priors}, we summarize the uniform (flat) priors adopted in our statistical analyses. These choices are motivated by the discussions presented in the previous sections and are selected to be sufficiently broad so as not to bias the inferred constraints. 

We close this section describing two statistical measures which are frequently used to judge the observational fitness of the extended $w_{\dagger}$VCDM  models with respect to a reference model.  
We consider two distinct statistical measures, namely, $\chi^2_{\rm min}$ and Akaike Information Criterion (AIC).\footnote{The AIC is defined as \cite{1100705},  $\text{AIC} = -2 \ln \mathcal{L}_{\text{max}} + 2n_0$,  where $\mathcal{L}_{\text{max}}$ corresponds to the maximum likelihood obtained for the model considering a specific observational dataset and $n_0$ is the number of free parameters of the model.} We compute \(\Delta \chi^2_{\rm min}\), defined as
\begin{equation}
\Delta \chi^2_{\rm min}
=
\chi^2_{\rm min}(\text{our model})
-
\chi^2_{\rm min}(\text{ref}),
\end{equation}
and \(\Delta\mathrm{AIC}\), defined as
\begin{equation}
\Delta\mathrm{AIC}
=
\mathrm{AIC}_{\rm min}(\text{our model})
-
\mathrm{AIC}_{\rm min}(\text{ref}).
\end{equation}

In both cases, a reference model must be specified. Since the cosmological models considered here are extended versions of the \(w_{\dagger}\)VCDM model that include the neutrino sector, the reference model must also be augmented with the same neutrino parameters; otherwise, a meaningful comparison cannot be performed.

Accordingly, the reference model ``ref'' is chosen to be the \(\Lambda\)CDM model supplemented with the same neutrino-sector parameters as those allowed to vary in the model under consideration (i.e., \(\Lambda\)CDM\(+\sum m_\nu\) when \(\sum m_\nu\) is free; \(\Lambda\)CDM\(+\sum m_\nu+N_{\rm eff}\) when both \(\sum m_\nu\) and \(N_{\rm eff}\) are free; and \(\Lambda\)CDM\(+m_{\rm sterile}\) when \(m_{\rm sterile}\) is free). In this way, the information criteria penalize only the additional dark-energy parameters associated with the \(w_{\dagger}\)VCDM sector.

Negative values of \(\Delta \chi^2_{\rm min}\) and \(\Delta\mathrm{AIC}\) indicate a better fit of the model relative to the corresponding \(\Lambda\)CDM-based reference.\footnote{According to the definitions of $\chi^2_{\rm min}$ and the AIC, these two quantities differ by twice the number of free parameters of the model.}
 In what follows, we discuss our main results.

\begin{table}
	\begin{center}
		\renewcommand{\arraystretch}{1.4}
		\begin{tabular}{|c@{\hspace{1 cm}}|@{\hspace{1 cm}} c|}
			\hline
			\textbf{Parameter}           & \textbf{Prior}\\
			\hline\hline

            $\omega_{b}$                 & $[0.0,\,1.0]$ \\ 
$\omega_{\rm cdm}$           & $[0.0,\,1.0]$ \\ 
$\tau_{\rm reio}$            & $[0.004,\,0.8]$ \\ 
$n_s$                        & $[0.1,\,2.0]$ \\ 
$\ln(10^{10} A_s)$           & $[1.0,\,5.0]$ \\ 
$100\,\theta_s$              & $[0.5,\,2.0]$ \\ \hline
$\Delta$                     & $[-5,\,5]$ \\ 
$z_{\dagger}$                & $[0,\,10]$ \\ \hline
$\sum m_\nu\;[\mathrm{eV}]$  & $[0.0,\,10]$ \\ 
$N_{\rm eff}$                & $[1.0,\,5.0]$ \\ 
$m_s\;[\mathrm{eV}]$         & $[0.0,\,10]$ \\
			
			\hline
		\end{tabular}
	\end{center}
	\caption{Flat priors are imposed on the free cosmological and model parameters used in the statistical analyses. The neutrino-sector parameters are varied only in the extended scenarios discussed in the text. Here, \(\omega_b = \Omega_{\rm b} h^2\) denotes the physical baryon density, \(\omega_c = \Omega_{\rm c} h^2\) the physical cold dark matter density, \(\tau_{\rm reio}\) the optical depth to reionization, \(\theta_{\mathrm{s}}\) the angular scale of the sound horizon at recombination, \(A_{\mathrm{s}}\) the amplitude of the primordial scalar perturbations, and \(n_{\mathrm{s}}\) the scalar spectral index. The parameters \(\Delta\) and \(z_{\dag}\) are associated with the model under consideration in this work. The neutrino parameters are discussed in the main text.}
	\label{tab:priors}
\end{table}

\begin{table*}[htpb!]
\begin{center}
\caption{
Marginalized constraints, along with the mean values at 68\% CL, for both the free and some derived parameters of the models considered in this work, based on the CMB dataset and its combinations with DESI, PantheonPlus (PP), and Union3. 
All quoted constraints correspond to 68\% confidence level, except for the neutrino mass sum $\sum m_\nu$, for which we report 95\% CL upper limits. 
In the last rows, we present $\Delta \chi^2_{\text{min}}$ and $\Delta \text{AIC}$ for each dataset combination.}

\label{tab:planck+bao}
\renewcommand{\arraystretch}{1.2}
\resizebox{\textwidth}{!}{
\begin{tabular}{l||c|c|c} 
\hline
\textbf{Dataset} & \textbf{Planck+DR2} & \textbf{Planck+DR2+PP} & \textbf{Planck+DR2+Union3} \\
\hline
\textbf{Model} & \textbf{$w_\dagger$VCDM + $\sum m_{\nu}$ } & \textbf{$w_\dagger$VCDM + $\sum m_{\nu}$} & \textbf{$w_\dagger$VCDM + $\sum m_{\nu}$ } \\
& \textcolor{blue}{\textbf{$w_\dagger$VCDM + $\sum m_{\nu}$ + $N_{\rm eff}$}} 
& \textcolor{blue}{\textbf{$w_\dagger$VCDM + $\sum m_{\nu}$ + $N_{\rm eff}$}} 
& \textcolor{blue}{\textbf{$w_\dagger$VCDM + $\sum m_{\nu}$ + $N_{\rm eff}$}} \\
\hline\hline

$10^{2}\omega{}_{b}$ 
  & $2.242^{+0.012}_{-0.014}$ 
  & $2.246^{+0.013}_{-0.013}$ 
  & $2.244^{+0.013}_{-0.012}$ \\
& \textcolor{blue}{$2.229^{+0.017}_{-0.016}$} 
& \textcolor{blue}{$2.252^{+0.014}_{-0.013}$} 
& \textcolor{blue}{$2.237^{+0.015}_{-0.019}$} \\[0.1cm]

$\omega{}_{cdm}$ 
  & $0.11947^{+0.00079}_{-0.00077}$ 
  & $0.11880^{+0.00061}_{-0.00060}$ 
  & $0.11907^{+0.00065}_{-0.00066}$ \\
& \textcolor{blue}{$0.1169^{+0.0025}_{-0.0024}$} 
& \textcolor{blue}{$0.1202^{+0.0024}_{-0.0022}$} 
& \textcolor{blue}{$0.1172^{+0.0026}_{-0.0031}$} \\[0.1cm]

$100\theta{}_{s}$ 
  & $1.04192^{+0.00028}_{-0.00028}$ 
  & $1.04200^{+0.00028}_{-0.00028}$ 
  & $1.04199^{+0.00027}_{-0.00027}$ \\
& \textcolor{blue}{$1.04232^{+0.00044}_{-0.00045}$} 
& \textcolor{blue}{$1.04180^{+0.00040}_{-0.00040}$} 
& \textcolor{blue}{$1.04226^{+0.00050}_{-0.00051}$} \\[0.1cm]

$\ln10^{10}A_{s}$ 
  & $3.042^{+0.014}_{-0.016}$ 
  & $3.046^{+0.014}_{-0.014}$ 
  & $3.044^{+0.014}_{-0.014}$ \\
& \textcolor{blue}{$3.036^{+0.015}_{-0.017}$} 
& \textcolor{blue}{$3.049^{+0.017}_{-0.015}$} 
& \textcolor{blue}{$3.039^{+0.015}_{-0.016}$} \\[0.1cm]

$n_{s}$ 
  & $0.9667^{+0.0036}_{-0.0035}$ 
  & $0.9684^{+0.0032}_{-0.0033}$ 
  & $0.9675^{+0.0034}_{-0.0034}$ \\
& \textcolor{blue}{$0.9616^{+0.0058}_{-0.0066}$} 
& \textcolor{blue}{$0.9708^{+0.0050}_{-0.0046}$} 
& \textcolor{blue}{$0.9637^{+0.0060}_{-0.0068}$} \\[0.1cm]

$\tau_{\text{reio}}$ 
  & $0.0541^{+0.0071}_{-0.0080}$ 
  & $0.0564^{+0.0071}_{-0.0076}$ 
  & $0.0548^{+0.0071}_{-0.0073}$ \\
& \textcolor{blue}{$0.0544^{+0.0070}_{-0.0071}$} 
& \textcolor{blue}{$0.0560^{+0.0081}_{-0.0082}$} 
& \textcolor{blue}{$0.0552^{+0.0064}_{-0.0074}$} \\[0.1cm]

$\sum m_\nu\,[\mathrm{eV}]$ (95\% CL) 
  & $<0.135$ 
  & $<0.066$ 
  & $<0.101$ \\
& \textcolor{blue}{$<0.118$} 
& \textcolor{blue}{$<0.117$} 
& \textcolor{blue}{$<0.095$} \\[0.1cm]

$N_{\rm eff}$ 
  & $3.0328~(\mathrm{fixed})$ 
  & $3.0328~(\mathrm{fixed})$ 
  & $3.0328~(\mathrm{fixed})$ \\
& \textcolor{blue}{$2.874^{+0.141}_{-0.152}$} 
& \textcolor{blue}{$3.114^{+0.139}_{-0.128}$} 
& \textcolor{blue}{$2.916^{+0.154}_{-0.184}$} \\[0.1cm]

$\Delta$ 
  & $-0.273^{+0.065}_{-0.052}$ 
  & $-0.094^{+0.029}_{-0.025}$ 
  & $-0.157^{+0.041}_{-0.044}$ \\
& \textcolor{blue}{$-0.256^{+0.088}_{-0.078}$} 
& \textcolor{blue}{$-0.126^{+0.021}_{-0.023}$} 
& \textcolor{blue}{$-0.170^{+0.046}_{-0.043}$} \\[0.1cm]

$z_{\dagger}$ 
  & $0.535^{+0.037}_{-0.045}$ 
  & $0.508^{+0.093}_{-0.095}$ 
  & $0.514^{+0.074}_{-0.071}$ \\
& \textcolor{blue}{$0.513^{+0.047}_{-0.047}$} 
& \textcolor{blue}{$0.541^{+0.070}_{-0.082}$} 
& \textcolor{blue}{$0.510^{+0.063}_{-0.063}$} \\

\hline
$\mathrm{H}_0 \, [\mathrm{km/s/Mpc}]$ 
  & $64.59^{+0.96}_{-0.75}$ 
  & $67.36^{+0.46}_{-0.47}$ 
  & $66.42^{+0.66}_{-0.76}$ \\
& \textcolor{blue}{$64.22^{+1.56}_{-1.56}$} 
& \textcolor{blue}{$67.19^{+0.80}_{-0.63}$} 
& \textcolor{blue}{$65.66^{+1.04}_{-1.27}$} \\[0.1cm]

$\Omega_{\rm m}$ 
  & $0.3419^{+0.0084}_{-0.0112}$ 
  & $0.3120^{+0.0047}_{-0.0051}$ 
  & $0.3219^{+0.0075}_{-0.0075}$ \\
& \textcolor{blue}{$0.339^{+0.014}_{-0.015}$} 
& \textcolor{blue}{$0.3170^{+0.0044}_{-0.0043}$} 
& \textcolor{blue}{$0.3247^{+0.0076}_{-0.0076}$} \\[0.1cm]

$\mathrm{S}_{8}$ 
  & $0.839^{+0.019}_{-0.019}$ 
  & $0.827^{+0.010}_{-0.010}$ 
  & $0.831^{+0.015}_{-0.015}$ \\
& \textcolor{blue}{$0.835^{+0.012}_{-0.012}$} 
& \textcolor{blue}{$0.829^{+0.012}_{-0.010}$} 
& \textcolor{blue}{$0.829^{+0.011}_{-0.011}$} \\

\hline

$\Delta \chi^2_{\text{min}}$ 
  & $-10.22$ & $-10.50$ & $-14.02$ \\
& \textcolor{blue}{$-9.20$}  
& \textcolor{blue}{$-7.60$}  
& \textcolor{blue}{$-16.44$} \\

$\Delta{\text{AIC}}$ 
  & $-6.22$ & $-6.50$ & $-10.02$ \\
& \textcolor{blue}{$-5.20$} 
& \textcolor{blue}{$-3.60$}  
& \textcolor{blue}{$-12.44$} \\

\hline
\hline
\end{tabular}}
\end{center}
\end{table*}

\subsection{A note on the negative neutrino mass range}

Although neutrino oscillation experiments firmly establish that at least two neutrinos have non-zero masses (and hence that the physical sum satisfies $\sum m_\nu > 0$; see e.g.~\cite{Lesgourgues:2006nd,Lesgourgues:2014zoa,TopicalConvenersKNAbazajianJECarlstromATLee:2013bxd,Dvorkin:2019jgs}), it is nevertheless useful to recall that some recent cosmological data combinations may exhibit a preference for $\sum m_\nu < 0$ when the parameter space is extended to allow for a signed neutrino-mass parameter. 
A representative example is the signed-$\sum m_\nu$ analysis of Ref.~\cite{Green:2024xbb}, which finds that specific combinations of Planck CMB, ACT DR6 data (including CMB lensing), and BAO measurements from SDSS/eBOSS DR16 together with early DESI BAO results can show an apparent $\mathcal{O}(2\text{--}3\sigma)$ pull toward negative values.

Importantly, this behavior should not be interpreted as evidence for negative neutrino masses. 
Rather, the negative-$\sum m_\nu$ region provides a continuous phenomenological parametrization of departures from the standard neutrino free-streaming effect, effectively mimicking an enhanced lensing or clustering amplitude. This interpretation is reinforced by subsequent work showing that the preference for $\sum m_\nu\le 0$ closely tracks the lensing-amplitude/clustering-amplitude anomalies and depends sensitively on dataset choices and modeling assumptions. For instance, Refs.~\cite{Elbers:2024sha,Naredo-Tuero:2024sgf}
emphasize that the signed-$\sum m_\nu$ preference is best viewed as a tension diagnostic (particularly in the lensing sector) rather than a physical determination of the neutrino-mass sign. Other complementary studies obtain bounds that are statistically compatible with positive bounds $\sum m_\nu $~\cite{Ivanov:2026dvl,Jhaveri:2025neg,Giare:2025ath, Lynch:2025ine,DESI:2025ffm}. Moreover, as BAO measurements evolve, the inferred preference can shift: recent DESI DR2 analyses provide tight constraints on late-time distances and, within $\Lambda$CDM (and simple extensions), yield strong upper limits on $\sum m_\nu$ that are fully compatible with the physical domain \cite{DESI:2025zgx}
while dedicated studies combining DESI DR2 with CMB/lensing data find that any support for negative values becomes reduced and model-dependent, often pushing the posterior toward values close to the oscillation-implied minimum \cite{Elbers:2025vlz}. 

While the debate on this topic remains active, for the purpose of parameter inference and quoted limits we impose the physical prior $\sum m_\nu \geq 0$. This choice is also consistent with direct kinematic searches, such as KATRIN, which constrain the absolute neutrino-mass scale through tritium $\beta$-decay~\cite{KATRIN:2025Science}. In this spirit, signed-$\sum m_\nu$ fits are useful as a diagnostic of residual dataset tensions (notably in the lensing and clustering amplitudes), whereas the neutrino-mass constraints reported in this work are obtained by integrating the posterior only over the physical domain. With this convention, any apparent formal preference for $\sum m_\nu < 0$ does not bias the reported upper limits and should not be interpreted as evidence against massive neutrinos.

\section{Main Results}
\label{results}

Table~\ref{tab:planck+bao} presents the marginalized constraints for the $w_{\dagger}$VCDM framework, including its extensions with a free total neutrino mass $\sum m_\nu$ and with both $\sum m_\nu$ and $N_{\rm eff}$ as additional degrees of freedom. Results are shown for four dataset combinations, all based on \textit{Planck} CMB data and DESI DR2 BAO measurements, supplemented by PP and Union3. For all dataset combinations, the $w_{\dagger}$VCDM scenario yields a systematically improved fit relative to the corresponding $\Lambda$CDM reference model with the same neutrino-sector assumptions, while remaining fully consistent with constraints from standard cosmological observables.

We first consider the standard cosmological parameters. The baryon density, $100\,\omega_b$, remains tightly constrained across all datasets, exhibiting only mild shifts when $N_{\rm eff}$ is allowed to vary. This behavior is expected, as $\omega_b$ correlates weakly with $N_{\rm eff}$ through the CMB damping tail. Similarly, the cold dark matter density $\omega_{\rm cdm}$ stays close to its $\Lambda$CDM value in all cases, with a slight preference for lower values when $N_{\rm eff}$ is freed, reflecting the known degeneracy between the radiation content and the matter density that sets the acoustic scale.

The parameters describing the primordial power spectrum, namely $\ln(10^{10}A_s)$, $n_s$, and the reionization optical depth $\tau_{\rm reio}$, are fully consistent with the \textit{Planck} baseline expectations.  Dataset combination including PP shows marginal shifts toward slightly higher values of $A_s$ and $\tau_{\rm reio}$, but these remain well within the $1\sigma$ level. As anticipated, allowing $N_{\rm eff}$ to vary leads to a moderate broadening of the constraints on $n_s$ and $A_s$, reflecting the well-known degeneracy between the relativistic energy density and the primordial spectral tilt.

Turning to the neutrino sector, we find that the $w_{\dagger}$VCDM dynamics preserves—and in some cases mildly strengthens—the stringent cosmological limits on the total neutrino mass. The 95\% CL upper bounds on $\sum m_\nu$ range from $\sum m_\nu < 0.066\,\mathrm{eV}$ for the Planck+DESI DR2+PP combination to $\sum m_\nu < 0.135\,\mathrm{eV}$ for Planck+DESI DR2 alone, remaining fully competitive with, and in several cases comparable to, those obtained within $\Lambda$CDM. When $N_{\rm eff}$ is treated as a free parameter, the bounds on $\sum m_\nu$ are slightly relaxed, as expected from parameter degeneracies, but remain robustly below the $0.12\,\mathrm{eV}$ level.

The inferred values of the effective number of relativistic species, $N_{\rm eff}$, are consistent with the standard value $N_{\rm eff}=3.046$ within $1\sigma$ for all dataset combinations. While the inclusion of PP data shows a mild preference for $N_{\rm eff}>3$, and other datasets weakly favor lower values, none of these shifts is statistically significant, providing no compelling evidence for additional relativistic degrees of freedom beyond the Standard Model.

\begin{figure*}[htpb!]
    \centering
    \includegraphics[width=0.9\textwidth]{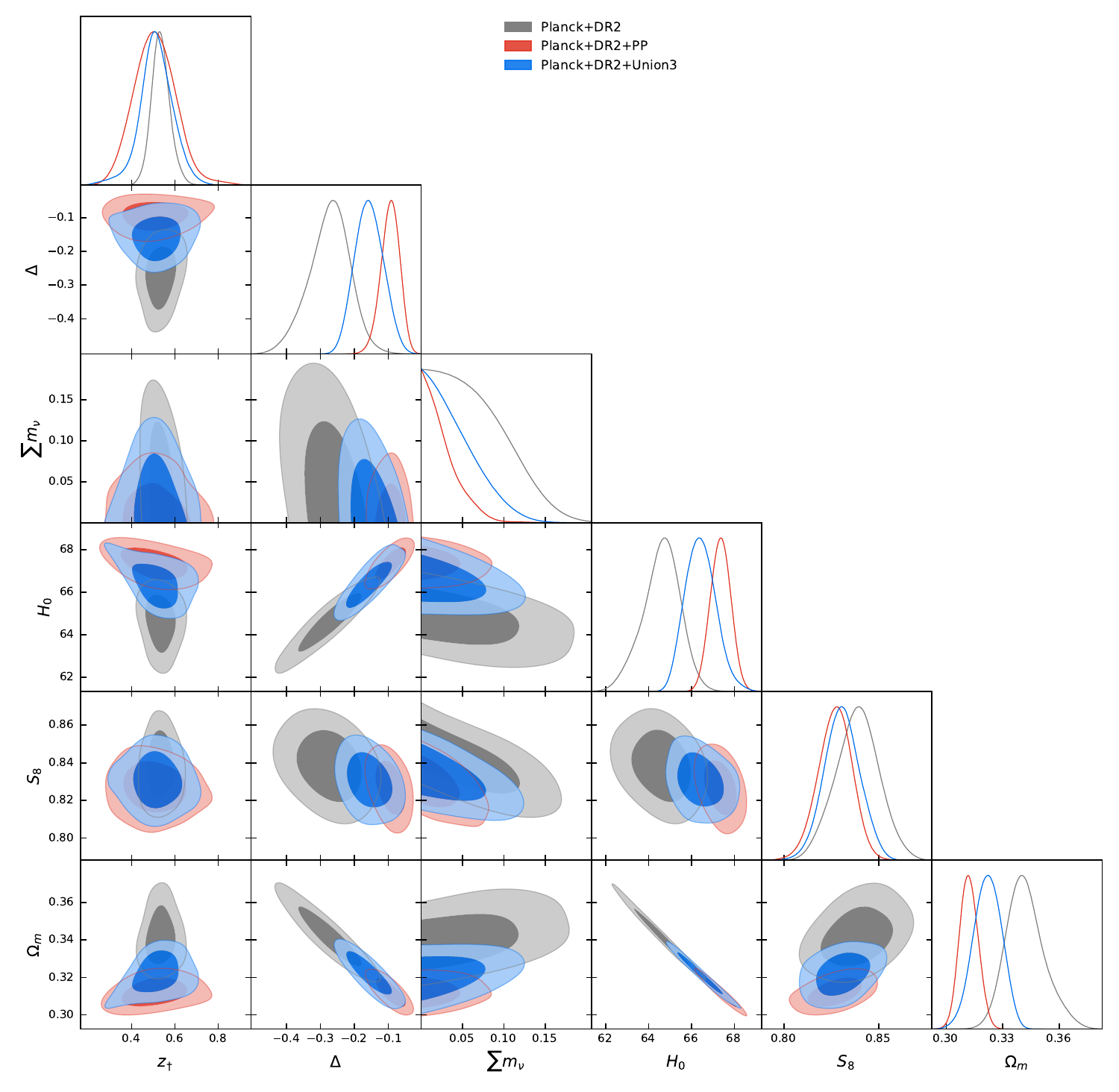}
    \caption{One-dimensional posterior distributions and 68\% and 95\% CL joint contours representing the $w_\dagger$VCDM + $\sum m_{\nu}$ scenario for several combined datasets. }
    \label{PS_model_results-1}
\end{figure*}

\begin{figure*}[htpb!]
    \centering
    \includegraphics[width=0.9\textwidth]{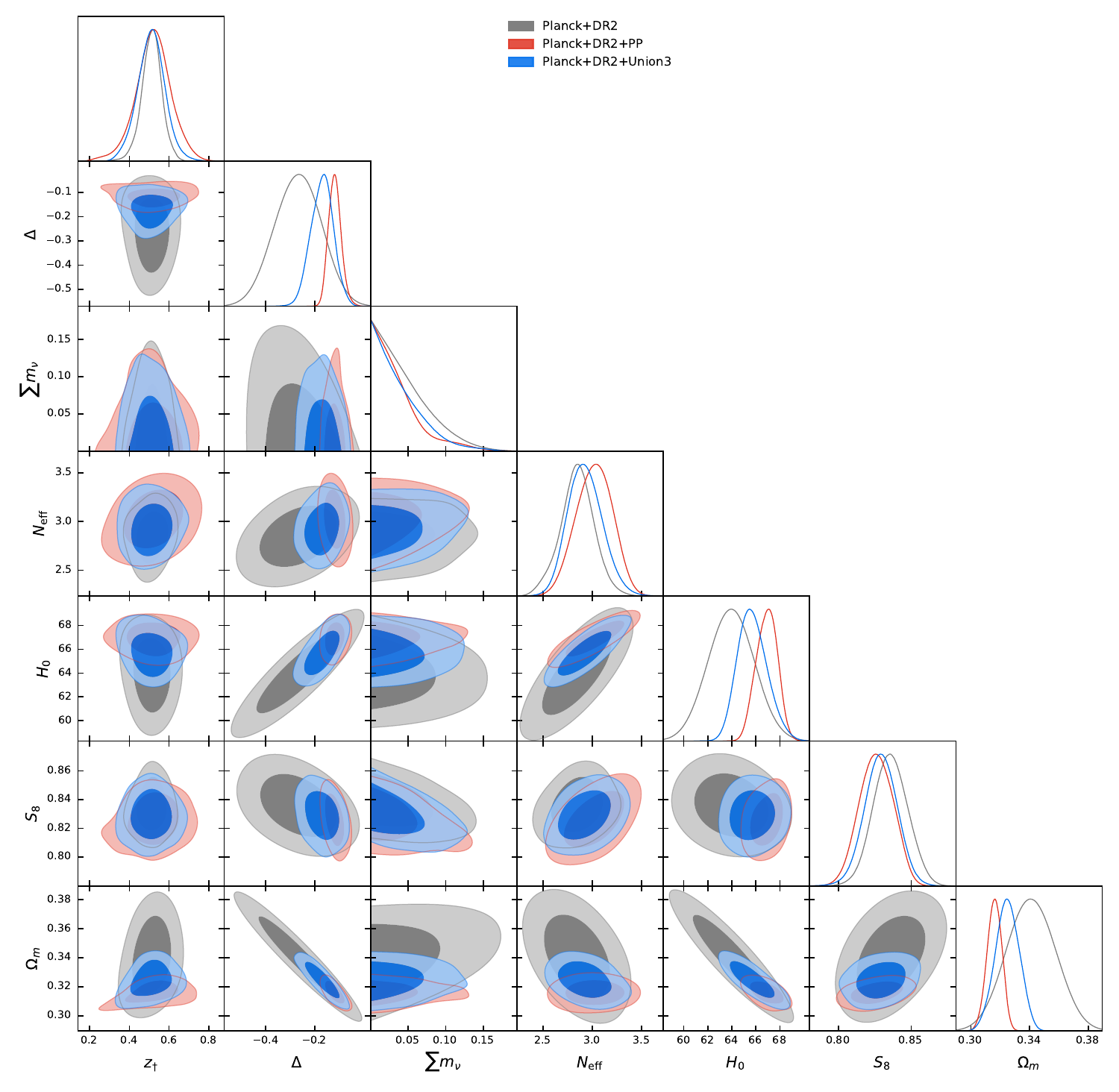}
    \caption{One-dimensional posterior distributions and 68\% and 95\% CL joint contours representing the $w_\dagger$VCDM + $\sum m_{\nu}$ + $N_{\rm eff}$ scenario for several combined datasets.}
    \label{PS_model_results-2}
\end{figure*}

The DE transition parameters, $\Delta$ and $z_{\dagger}$, encode the central physical features of the $w_{\dagger}$VCDM scenario. Across all dataset combinations, we find a clear and robust preference for $\Delta<0$, indicating a transition from a phantom-like regime ($w<-1$) at earlier times to a quintessence-like EoS ($w>-1$) at late times. This result is particularly noteworthy, as it implies that the present-day Universe is consistently described by non-phantom DE, while allowing for phantom behavior in the past, fully driven by observational data rather than theoretical assumptions. 

To quantify the level of agreement in a statistically consistent manner, we adopt the quadratic ``parameter-shift'' estimator introduced in~\cite{Addison:2015wyg, ACT:2020gnv} (see also~\cite{Nunes:2021ipq}). We stress that the validity of this estimator hinges on a correct treatment of the covariance of the difference between parameter estimates, in particular on properly accounting for statistical dependence whenever the datasets being compared share common information \cite{RaveriHu:2018, RaveriZacharegkasHu:2019}.

Let us consider two datasets, labeled $i$ and $j$, yielding mean parameter vectors $\mathbf{x}_i$ and $\mathbf{x}_j$ with associated covariance matrices $C_i$ and $C_j$. The level of agreement is assessed through the difference vector
\begin{equation}
\Delta\mathbf{x} \equiv \mathbf{x}_i - \mathbf{x}_j .
\end{equation}
In full generality, the covariance of this difference is
\begin{equation}
\mathrm{Cov}(\Delta\mathbf{x}) = C_i + C_j - 2\,\mathrm{Cov}(\mathbf{x}_i,\mathbf{x}_j).
\end{equation}
Only if the two datasets are statistically independent does the cross-covariance term vanish, reducing the expression to $C_i+C_j$. This simplification is not justified when the datasets share common information (e.g., when one dataset is nested within the other), in which case retaining the cross-covariance is essential to avoid misestimating the variance of $\Delta\mathbf{x}$ and hence biasing the inferred significance of apparent parameter shifts \cite{RaveriHu:2018, RaveriZacharegkasHu:2019}.

For nested datasets, where dataset $j$ contains all the information of dataset $i$ plus additional (statistically independent) data, the cross-covariance is well approximated by $\mathrm{Cov}(\mathbf{x}_i,\mathbf{x}_j)\simeq C_j$, which leads to $\mathrm{Cov}(\Delta\mathbf{x})\simeq C_i-C_j$. This makes explicit that the shift between the two estimates is driven by the additional information in dataset $j$, rather than by the information common to both.

The quadratic statistic is therefore defined as
\begin{equation}
\chi^2 =
(\mathbf{x}_i - \mathbf{x}_j)^{\mathrm T}
\left[ C_i + C_j - 2\,\mathrm{Cov}(\mathbf{x}_i,\mathbf{x}_j) \right]^{-1}
(\mathbf{x}_i - \mathbf{x}_j),
\label{eq:tension_stat}
\end{equation}
which is the quadratic form associated with the covariance of the difference vector, i.e., a generalized Mahalanobis distance.

Interpreting Eq.~\eqref{eq:tension_stat} in terms of a $\chi^2$ distribution and translating it into a Gaussian-equivalent significance relies on the assumption that $\Delta\mathbf{x}$ is approximately multivariate normal with the covariance defined above. Following standard practice, we compute the probability-to-exceed as a tail probability, ${\rm PTE}\equiv P(\chi^2_k\ge \chi^2_{\rm obs})$, and quote the corresponding two-sided Gaussian-equivalent significance as $Z\equiv \Phi^{-1}(1-{\rm PTE}/2)$ \cite{Cowan:2010}. We note that alternative one-dimensional proxy mappings based on an error-function inversion (often denoted $N_\sigma$) also appear in the literature \cite{Lemos:2020}, but here we adopt the tail-PTE $\rightarrow Z$ convention to match the usual interpretation of a $\chi^2$ test.

In our analysis, Eq.~\eqref{eq:tension_stat} is evaluated in the full parameter space, using $k=9$ parameters in the baseline model and $k=10$ when allowing $N_{\rm eff}$ to vary. For each dataset combination, the mean parameter vectors and covariance matrices are estimated directly from the MCMC chains. For nested comparisons of the form (\textit{Planck}+DR2) versus (\textit{Planck}+DR2+X), where X run over others datasets, we adopt the nested-dataset approximation $\mathrm{Cov}(\mathbf{x}_i,\mathbf{x}_j)\simeq C_j$, such that $\mathrm{Cov}(\Delta\mathbf{x})\simeq C_i-C_j$. With this statistically consistent treatment, we find that the largest shift arises when adding PantheonPlus: in the baseline model we obtain $Z\simeq 2.64\sigma$ for (\textit{Planck}+DR2) versus (\textit{Planck}+DR2+PantheonPlus), while (\textit{Planck}+DR2) versus (\textit{Planck}+DR2+Union3) yields $Z\simeq 0.89\sigma$. When allowing $N_{\rm eff}$ to vary, the shifts decrease to $Z\simeq 0.59\sigma$ (PantheonPlus) and $Z\simeq 0.06\sigma$ (Union3), as expected when the additional degree of freedom enlarges the available degeneracy directions and can absorb part of the shift induced by the extra dataset. Finally, for the overlap comparison (\textit{Planck}+DR2+PantheonPlus) versus (\textit{Planck}+DR2+Union3), the datasets are not nested but share a common subset; in this case the cross-covariance does not vanish and the tension remains essentially null, with $Z\approx 0$ in the baseline model and $Z\simeq 0.013\sigma$ when allowing $N_{\rm eff}$, consistent with the expectation that small differences at the $\sim10^{-2}$ level are driven by Monte Carlo noise in the estimated covariances and by the modest reshuffling of degeneracy directions when increasing $k$ by one.

The corner plot shown in Fig.~\ref{PS_model_results-1} reveals that the posteriors obtained from the different dataset combinations occupy distinct but overlapping regions of parameter space, indicating that the observed shifts are driven by correlated parameter adjustments rather than by internal inconsistencies among the data. It is important to emphasize, as documented in the recent literature, that the DESI DR2 BAO data and the SNIa samples tend to pull the matter density parameter, $\Omega_m$, in opposite directions e.g., ~\cite{Bousis:2024rnb,Pedrotti:2024kpn,DES:2025bxy}. This mild tension propagates to other cosmological parameters through their statistical correlations, inducing, for instance, a strong anti-correlation in the $\Omega_m$--$\Delta$ parameter plane.
 
A particularly relevant feature of the analysis is the presence of a clear, albeit mild, anti-correlation between $\Delta$ and the total neutrino mass, $\sum m_\nu$. More negative values of $\Delta$, corresponding to a stronger phantom-to-quintessence transition, can be partially compensated by larger neutrino masses, which suppress the growth of cosmic structures on small scales. As a consequence, dataset combinations that allow or favor higher values of $\sum m_\nu$ tend to shift the posterior distribution of $\Delta$ toward more negative values, whereas tighter constraints on neutrino masses significantly reduce the allowed range of $\Delta$. The inclusion of SNIa data plays a crucial role in shaping these parameter degeneracies. In particular, the PantheonPlus compilation provides stronger constraints on the late-time expansion history, effectively reducing the available parameter volume along the $\Delta$ direction. This results in a noticeable shift of the Planck+DR2+PantheonPlus posterior toward less negative values of $\Delta$ when compared to the Planck+DR2 case. In contrast, the Union3 compilation is slightly more permissive over the relevant redshift range, allowing for a broader overlap with the Planck+DR2 constraints and leading to a higher degree of mutual compatibility between the corresponding dataset combinations.
 
Additional insight is provided by the correlations involving the Hubble constant. The corner plot~\ref{PS_model_results-1} shows the expected anticorrelation between $\sum m_\nu$ and $H_0$, whereby larger neutrino masses drive the inferred value of $H_0$ toward lower values. The inclusion of SNIa data shifts the preferred region of $H_0$ to higher values, which in turn disfavors large neutrino masses and indirectly constrains the allowed range of $\Delta$. This three-parameter interplay between $\Delta$, $\sum m_\nu$, and $H_0$ explains the moderate shifts observed between the different dataset combinations. The impact of $\sum m_\nu$ is also evident in the correlations with structure-growth parameters such as $S_8$ and $\Omega_m$. Allowing $\sum m_\nu$ to vary broadens the posteriors in the $S_8$ direction, reflecting the well-known suppression of structure formation induced by massive neutrinos. The correlations with $\Omega_m$ remain comparatively mild, indicating that the dominant degeneracy directions involve $\Delta$, $\sum m_\nu$, $H_0$, and $S_8$, rather than $\Omega_m$ alone.
 
Overall, the parameter shifts observed across the different dataset combinations are naturally explained by the physical degeneracies introduced by massive neutrinos and by the complementary constraining power of late-time distance indicators. The results therefore point to a coherent picture in which the datasets are mutually consistent, and the observed differences arise from correlations in the extended parameter space. The same interpretation discussed above also applies in the context of the $w_\dagger$VCDM$+\sum m_\nu + N_{\rm eff}$ model, once the additional parameter correlations induced by $N_{\rm eff}$ are properly taken into account.

The other main parameter that deserves careful interpretation is $z_{\dagger}$. 
As shown in \cite{Scherer:2025esj}, the error bars and the inferred accuracy of $z_{\dagger}$ remain robust across different joint analyses. 
Here, it is again important to note that we introduce new parameters, such as $\sum m_\nu$ and $N_{\rm eff}$, which generate additional correlations with key cosmological parameters like $H_0$ and $\Omega_m$. 
In turn, these correlations can directly affect the inferred constraints on $z_{\dagger}$. At first sight, it may appear counterintuitive that the inclusion of additional datasets, such as Type~Ia supernovae, leads to broader marginalized uncertainties on $z_{\dagger}$. 
However, this behavior is a natural consequence of the multidimensional structure of the parameter space and of the degeneracies that arise when extending the model to include free neutrino parameters.

The increase in the marginalized uncertainty on $z_{\dagger}$ when Union3 or PantheonPlus supernova data are added to the Planck+DESI~DR2 baseline is a outcome of the physical and statistical structure of the parameter space. Supernova data strongly constrain the late-time expansion, favoring a smaller transition amplitude $|\Delta|$, which in turn makes the DE equation-of-state evolution smoother and less sensitive to the exact transition redshift. At the same time, degeneracies between $z_{\dagger}$, $\Delta$, $\sum m_\nu$ and $H_0$, especially when $\sum m_\nu$ is free, propagate additional uncertainty into the posterior of $z_{\dagger}$ through marginalization, further broadening its constraints when the transition is already gradual.

This behavior is therefore not an anomaly but reflects the complementary roles of different cosmological probes: while supernova data tightly pin down the late-time expansion and hence the amplitude $\Delta$, they provide comparatively weaker direct leverage on the epoch $z_{\dagger}$ of a smooth transition. The resulting increase in the uncertainty of $z_{\dagger}$ is coherent with the model's construction and highlights how parameter constraints can evolve in a correlated, multi-dimensional space when new datasets are incorporated. The same interpretation applies to the results derived within the $w_\dagger$VCDM$+\sum m_\nu + N_{\rm eff}$ model.


Several derived parameters also display systematic trends. The inferred Hubble constant increases when low-redshift datasets are included: while the CMB+DR2 combination alone yields $H_0\simeq 64.6\,\mathrm{km\,s^{-1}\,Mpc^{-1}}$, the addition of PP and Union3 shifts the preferred value to $H_0\simeq 66.4$--$67.4\,\mathrm{km\,s^{-1}\,Mpc^{-1}}$. Although this does not fully resolve the Hubble tension, the shift occurs in the correct direction and suggests that the $w_{\dagger}$ transition partially alleviates the discrepancy. The matter density parameter is consistently constrained to $\Omega_{\rm m}\simeq 0.31$--$0.34$, with a mild reduction when PP data are included. The clustering parameter $S_8$ remains in the range $S_8\simeq 0.826$--$0.839$, fully consistent with \textit{Planck} constraints and only marginally closer to weak-lensing preferred values, indicating that the model does not significantly modify the current level of the $S_8$ tension.

\begin{figure*}
    \centering
\includegraphics[width=1.0\textwidth]{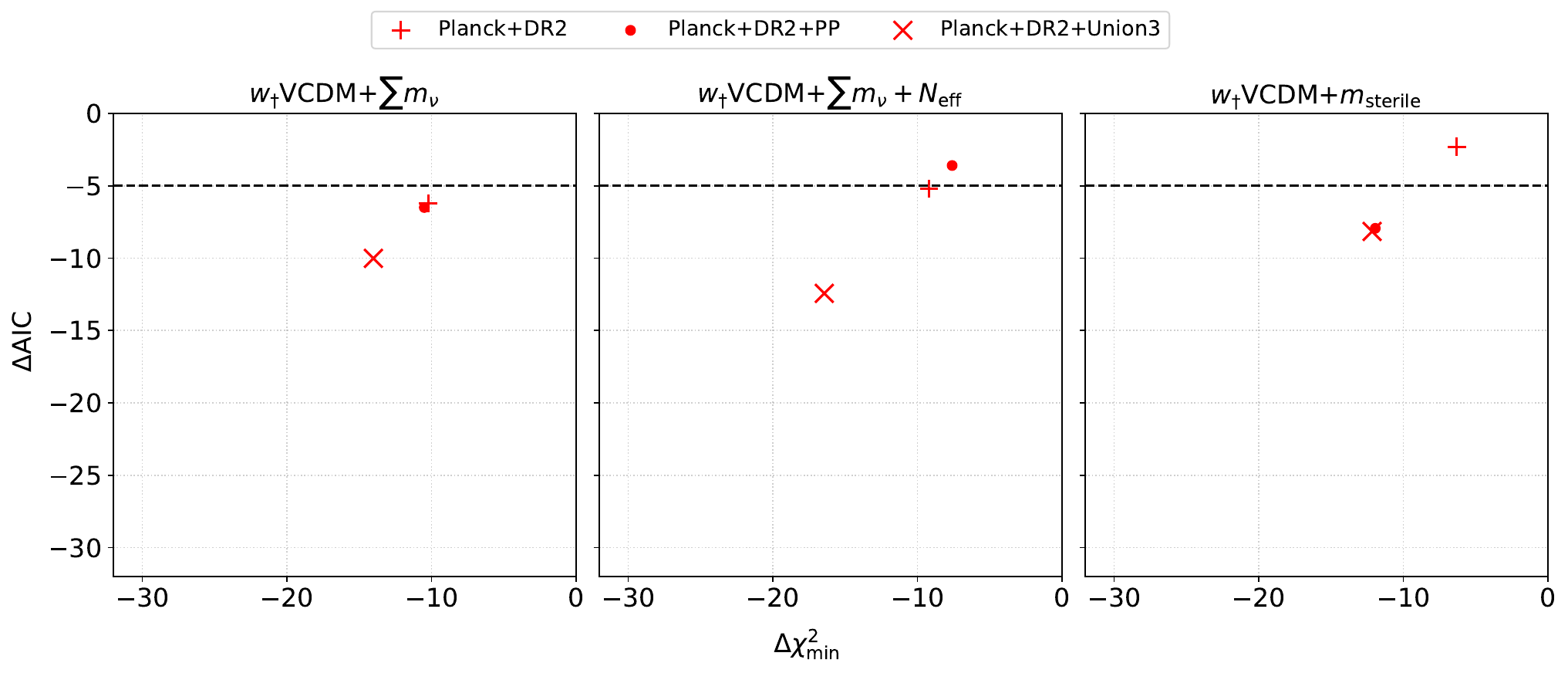}

    \caption{The figure shows AIC versus $\Delta \chi^2_{\rm min}$ for the extended $w_{\dagger}$VDM models considering all the datasets. }
    \label{fig:AIC}
\end{figure*}

Finally, the goodness-of-fit statistics provide strong quantitative support for the $w_{\dagger}$VCDM framework. For all dataset combinations,  $\chi^2_{\rm min}$ for the model is reduced relative to the corresponding $\Lambda$CDM reference model, with $\Delta\chi^2_{\rm min}$ ranging from approximately $-11$ to $-29$. Crucially, even after accounting for the additional DE parameters through the Akaike Information Criterion, all combinations yield negative values of $\Delta\mathrm{AIC}$. This confirms that the improvement in fit is not driven by overfitting but is statistically meaningful. In Fig. \ref{fig:AIC} we show the graphical variations between $\Delta\chi^2_{\rm min}$ and $\Delta\mathrm{AIC}$ for all the datasets which clearly shows the preference of \textbf{$w_\dagger$VCDM + $\sum m_{\nu}$ } and \textbf{$w_\dagger$VCDM + $\sum m_{\nu}$ + $N_{\rm eff}$} over the standard $\Lambda$CDM model in the equivalent neutrino framework. 
This is the central outocme  of the current article since the evidence of dynamical DE remains robust in presence of the neutrino sector. 

\begin{figure*}[htpb!]
    \centering
    \includegraphics[width=0.9\textwidth]{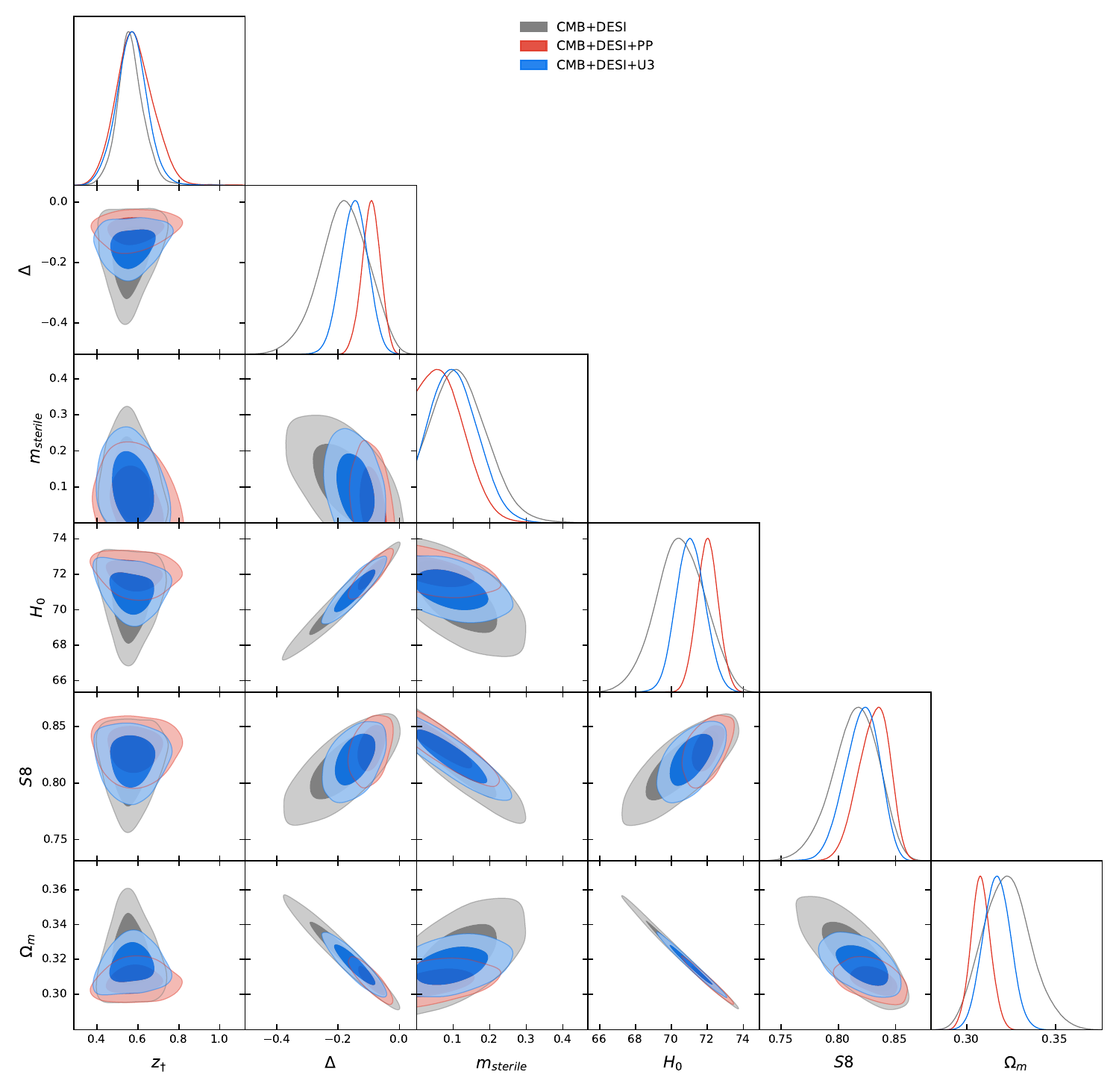}
    \caption{One-dimensional posterior distributions and 68\% and 95\% CL joint contours representing the $w_\dagger$VCDM + $m_{esterile}$ scenario for several combined datasets.}
    \label{PS_model_results-2}
\end{figure*}

\begin{table*}[htpb!]
\begin{center}
\caption{
Marginalized constraints and posterior mean values at the $68\%$ confidence level are reported for both the free and selected derived parameters of the $w_\dagger$VCDM+$m_{\rm sterile}$ model, using \textit{Planck} CMB data in combination with DESI DR2 and with the additional inclusion of PantheonPlus (PP) or Union3.
Unless explicitly stated otherwise, all quoted uncertainties correspond to $68\%$ confidence intervals.
In the last rows, we present $\Delta \chi^2_{\text{min}}$ and $\Delta \text{AIC}$ for each dataset combination.}

\label{tab:planck+bao_esteril}
\renewcommand{\arraystretch}{1.2}
\resizebox{\textwidth}{!}{
\begin{tabular}{l||c|c|c} 
\hline
\textbf{Dataset} & \textbf{Planck+DR2} & \textbf{Planck+DR2+PP} & \textbf{Planck+DR2+Union3} \\
\hline
\textbf{Model} & \multicolumn{3}{c}{\textbf{$w_\dagger$VCDM + $m_{\rm sterile}$ (sterile neutrino)}} \\
\hline\hline

$10^{2}\omega{}_{b}$ 
  & $2.310^{+0.013}_{-0.014}$ 
  & $2.311^{+0.012}_{-0.014}$ 
  & $2.310^{+0.013}_{-0.013}$ \\[0.1cm]

$\omega{}_{cdm}$ 
  & $0.13511^{+0.00099}_{-0.00100}$ 
  & $0.13490^{+0.00106}_{-0.00077}$ 
  & $0.13507^{+0.00103}_{-0.00092}$ \\[0.1cm]

$100\theta{}_{s}$ 
  & $1.03994^{+0.00028}_{-0.00028}$ 
  & $1.03993^{+0.00029}_{-0.00028}$ 
  & $1.03994^{+0.00028}_{-0.00028}$ \\[0.1cm]

$\ln10^{10}A_{s}$ 
  & $3.092^{+0.015}_{-0.017}$ 
  & $3.092^{+0.014}_{-0.017}$ 
  & $3.091^{+0.016}_{-0.016}$ \\[0.1cm]
  
$n_{s}$ 
  & $0.9993^{+0.0036}_{-0.0037}$ 
  & $0.9997^{+0.0037}_{-0.0033}$ 
  & $0.9993^{+0.0036}_{-0.0036}$ \\[0.1cm]

$\tau_{\text{reio}}$ 
  & $0.0618^{+0.0076}_{-0.0091}$ 
  & $0.0617^{+0.0074}_{-0.0090}$ 
  & $0.0611^{+0.0077}_{-0.0088}$ \\[0.1cm]

$m_{\rm sterile}\,[\mathrm{eV}]$ (95\% CL) 
  & $<0.264$ 
  & $<0.185$ 
  & $<0.234$ \\[0.1cm]

$\Delta$ 
  & $-0.167^{+0.091}_{-0.073}$ 
  & $-0.092^{+0.032}_{-0.029}$ 
  & $-0.142^{+0.047}_{-0.045}$ \\[0.1cm]

$z_{\dagger}$ 
  & $0.562^{+0.071}_{-0.063}$ 
  & $0.586^{+0.083}_{-0.095}$ 
  & $0.565^{+0.075}_{-0.073}$ \\

\hline
$\mathrm{H}_0 \, [\mathrm{km/s/Mpc}]$ 
  & $70.76^{+1.52}_{-1.38}$ 
  & $72.01^{+0.58}_{-0.60}$ 
  & $71.17^{+0.78}_{-0.91}$ \\[0.1cm]

$\Omega_{\rm m}$ 
  & $0.3202^{+0.0128}_{-0.0155}$ 
  & $0.3077^{+0.0053}_{-0.0057}$ 
  & $0.3159^{+0.0084}_{-0.0078}$ \\[0.1cm]

$\mathrm{S}_{8}$ 
  & $0.844^{+0.014}_{-0.014}$ 
  & $0.841^{+0.014}_{-0.010}$ 
  & $0.843^{+0.014}_{-0.012}$ \\

\hline
$\Delta \chi^2_{\text{min}}$ 
  & $-6.30$   
  & $-11.94$  
  & $-12.16$  
  \\[0.1cm]

$\Delta{\text{AIC}}$ 
  & $-2.30$   
  & $-7.94$   
  & $-8.16$   
  \\

\hline
\hline
\end{tabular}}
\end{center}
\end{table*}

\subsection{Constraints in presence of Sterile Neutrino}
\label{sec:results_esterile}

In this section, we present the cosmological constraints obtained for the $w_\dagger$VCDM model in the presence of an additional sterile neutrino species. We assume the sterile neutrino to be fully thermalized, thereby extending the standard neutrino sector beyond the three active species. Constraints and upper limits on the mass of the sterile neutrino, as well as their associated uncertainties, within the \(\Lambda\)CDM model and its extensions to dynamical dark energy models have been investigated previously (see, for example, Ref.~\cite{Abazajian:2017tcc,Hu:2025lrl,Boyarsky:2009ix,Feng:2021ipq,Du:2025iow,Feng:2025wbz,Benso:2024qrg,Hagstotz:2020ukm}). Here, we present a generalization of these analyses to our \(w_\dagger\)VCDM scenario.

Following the prescription adopted in Ref.~\cite{Planck:2013pxb} 
and subsequent cosmological analyses, the sterile-neutrino contribution is incorporated by promoting the effective number of relativistic species, $N_{\rm eff}$, and introducing an effective sterile-neutrino mass parameter, $m^{\rm eff}_{\nu,\,\rm sterile}$. The latter is defined as
\begin{equation}
m^{\rm eff}_{\nu,\,\rm sterile} \equiv \left(\Delta N_{\rm eff}\right)^{3/4} m^{\rm thermal}_{\nu,\,\rm sterile},
\end{equation}
where $\Delta N_{\rm eff} \equiv N_{\rm eff} - N_{\rm SM}$, and $N_{\rm SM} = 3.044$ denotes the Standard Model prediction for the effective number of active neutrino species. The parameter $m^{\rm thermal}_{\nu,\,\rm sterile}$ corresponds to the physical mass of a thermally produced sterile neutrino.

In our analysis, we impose a conservative prior on the sterile-neutrino mass, $m^{\rm thermal}_{\nu,\,\rm sterile} < 10~\mathrm{eV}$, consistent with the assumptions adopted in Ref.~\cite{Planck:2018vyg}  and ensuring that the sterile component does not behave as cold dark matter on cosmological scales. This choice allows us to explore the phenomenological impact of a sterile neutrino while remaining compatible with current cosmological bounds.

For the active neutrino sector, we fix the total mass of the three active species to the minimal value $\sum m_\nu = 0.06~\mathrm{eV}$, assuming the normal mass ordering. This setup isolates the effects of the sterile neutrino and avoids degeneracies associated with simultaneously varying multiple neutrino mass parameters. Within this framework, we investigate how the presence of a sterile neutrino modifies the constraints on the DE transition parameters of the $w_\dagger$VCDM model, as well as its impact on key derived cosmological parameters such as the Hubble constant, matter density, and clustering amplitude.

Table~\ref{tab:planck+bao_esteril} summarizes the posterior constraints on all free and derived parameters of the $w_\dagger$VCDM+$m_{\rm sterile}$ model, obtained using \textit{Planck} data alone and in combination with DESI DR2, PantheonPlus (PP) and Union3.

Across all dataset combinations, the effective mass of the sterile neutrino is not detected at a statistically significant level. Instead, we obtain robust upper bounds at the $95\%$ confidence level,
\[
m_{\rm sterile} < 
\begin{cases}
0.264~\mathrm{eV} & (\text{Planck+DR2}),\\
0.185~\mathrm{eV} & (\text{Planck+DR2+PP}),\\
0.234~\mathrm{eV} & (\text{Planck+DR2+Union3}),\\
\end{cases}
\]

These limits indicate that current cosmological data remain fully compatible with the absence of a massive sterile neutrino, while still allowing for a light, fully thermalized state with sub-eV mass. The inclusion of low-redshift distance indicators, particularly PantheonPlus, significantly tightens the constraints, reflecting their sensitivity to late-time expansion history.

Importantly, constraints demonstrate that the dynamics of  $w_\dagger$VCDM effectively absorbs part of the phenomenology typically attributed to additional relativistic degrees of freedom, thus relaxing the need for a massive sterile component while maintaining consistency with all datasets.

The DE transition parameters, namely the transition amplitude $\Delta$ and the transition redshift $z_\dagger$, remain remarkably stable with respect to the inclusion of a sterile neutrino. We find consistent evidence for $\Delta<0$ across all dataset combinations, corresponding to a transition from a phantom-like EoS ($w<-1$) at high redshift to a quintessence-like regime ($w>-1$) at late times. The preferred values of $\Delta$ are mildly dataset dependent, with PantheonPlus again favoring smaller deviations from $\Lambda$CDM.

The transition redshift is tightly constrained,
\[
z_\dagger \simeq 0.56 \pm 0.07,
\]
with variations below the $5\%$ level across all datasets. This robustness confirms that the late-time nature of the transition is a genuine feature of the data rather than a consequence of the sterile neutrino sector.

Despite the absence of a significant detection of $m_{\rm sterile}$, the combined effect of the DE transition and the extended neutrino sector has a pronounced impact on derived parameters. In particular, the inferred value of the Hubble constant is significantly increased in the $w_\dagger$VCDM framework with an extended neutrino sector. Across all dataset combinations, we find
\[
H_0 \approx 71.0\text{--}72.0~\mathrm{km\,s^{-1}\,Mpc^{-1}},
\]
representing an upward shift of approximately $4$--$5~\mathrm{km\,s^{-1}\,Mpc^{-1}}$ relative to the \textit{Planck} $\Lambda$CDM baseline~\cite{Planck:2018vyg}. As a consequence, the tension with local distance-ladder measurements is substantially reduced.

This increase in $H_0$ can be physically understood as the result of a well-known degeneracy between the early-time radiation content and the late-time expansion rate (see, for example,~\cite{GarciaEscudero:2025orc,Escudero:2022rbq,Pan:2023frx,Dhuria:2023yrw} for previous discussions in this regard.). An increase in the effective number of relativistic species, $N_{\rm eff}$, enhances the expansion rate prior to recombination, thus reducing the sound horizon at decoupling. To preserve the observed angular scale of the acoustic peaks in the CMB, this effect is compensated by a higher value of $H_0$. In the present model, this mechanism operates in synergy with the late-time DE transition encoded by the parameters $\Delta$ and $z_\dagger$, which further modifies the history of the low-redshift expansion.

Overall, the combined effect of an enhanced radiation density at early times and a controlled modification of the late-time expansion history leads to a coherent upward shift in $H_0$, while remaining consistent with baryon acoustic oscillation and supernova constraints, reducing the Hubble tension from the $\sim 5\sigma$ level in $\Lambda$CDM to approximately the $2$--$2.5\sigma$ range, depending on the dataset combination. This highlights the $w_\dagger$VCDM+$m_{\rm sterile}$ scenario as a viable and phenomenologically compelling framework for addressing one of the most persistent discrepancies in modern cosmology.

Finally, regarding the statistical performance of the 
$w_\dagger$VCDM$+m_{\rm sterile}$ model, we find
\[
\Delta\chi^2_{\rm min} < 0 \qquad \text{and} \qquad \Delta\mathrm{AIC} < 0.
\]
These results demonstrate that the inclusion of a DE transition yields a significantly better fit to the data than $\Lambda$CDM, even when the sterile neutrino mass is constrained only through upper bounds. The improvement is therefore primarily driven by the modified late-time dynamics, with the sterile neutrino sector remaining a viable but subdominant extension. In the last panel of Fig. \ref{fig:AIC} we graphically demonstrate this behavior for all the datasets. 

\begin{table}[htpb!]
\centering
\caption{Difference in statistical significance between the free active-neutrino mass case and the sterile-neutrino mass case,
$\Delta\chi^2_{\min}\equiv \chi^2_{\min}(m_{\rm sterile}) - \chi^2_{\min}(\sum m_\nu\ {\rm free})$.}

\label{tab:dchi2_layout_compact}

\footnotesize
\setlength{\tabcolsep}{6pt}
\renewcommand{\arraystretch}{1.10}

\resizebox{\columnwidth}{!}{%
\begin{tabular}{l|c|c}
\hline\hline
\shortstack[l]{\Tstrut\textbf{$\Lambda$CDM}\\[-0.2ex]\textcolor{blue}{\textbf{$w_\dagger$VCDM}}\Bstrut}
& \makebox[2.6cm][c]{\Tstrut\textbf{$\Delta\chi^2_{\min}$}\Bstrut}
& \makebox[1.3cm][c]{\Tstrut\textbf{$\sigma$}\Bstrut} \\
\hline \hline

Planck+DR2
& \makebox[2.6cm][c]{$33.54$} & \makebox[1.3cm][c]{$5.79$} \\
& \textcolor{blue}{\makebox[2.6cm][c]{$37.46$}} & \textcolor{blue}{\makebox[1.3cm][c]{$6.12$}} \\
\hline

Planck+DR2+PP
& \makebox[2.6cm][c]{$38.86$} & \makebox[1.3cm][c]{$6.23$} \\
& \textcolor{blue}{\makebox[2.6cm][c]{$37.42$}} & \textcolor{blue}{\makebox[1.3cm][c]{$6.12$}} \\
\hline

Planck+DR2+Union3
& \makebox[2.6cm][c]{$31.18$} & \makebox[1.3cm][c]{$5.58$} \\
& \textcolor{blue}{\makebox[2.6cm][c]{$36.88$}} & \textcolor{blue}{\makebox[1.3cm][c]{$6.07$}} \\
\hline\hline
\end{tabular}}
\end{table}

On the other hand, it is important to emphasize that all comparisons presented in this section are performed with respect to the $\Lambda$CDM$+m_{\rm sterile}$ model. In particular, all quoted differences implicitly assume a fixed effective number of relativistic species, $N_{\rm eff}=4.04$, corresponding to the presence of one fully thermalized sterile neutrino.

A direct and informative comparison can therefore be made between analyses with and without the inclusion of a sterile neutrino mass. 
Table~IV summarizes the corresponding differences in goodness of fit through the quantity
\begin{equation}
\Delta\chi^2_{\min} \equiv 
\chi^2_{\min}(m_{\rm sterile}\ \mathrm{free}) 
- \chi^2_{\min}(\sum m_\nu\ \mathrm{free}) \, ,
\end{equation}
which quantifies the relative statistical preference between the sterile-neutrino scenario and the case in which only the sum of active neutrino masses is allowed to vary.

With this definition, positive values of $\Delta\chi^2_{\min}$ indicate that the model with a free sterile neutrino mass provides a poorer fit to the data, i.e. it is statistically disfavored with respect to the corresponding model with free active neutrino masses.
The second column of Table~IV reports the statistical significance of this exclusion, expressed in terms of the equivalent number of Gaussian standard deviations, $\sigma$.

Specifically, we show how strongly the hypotheses 
($\Lambda$CDM$+m_{\rm sterile}$) and ($w_\dagger$VCDM$+m_{\rm sterile}$) 
are disfavored relative to their counterparts 
($\Lambda$CDM$+\sum m_\nu$ free) and ($w_\dagger$VCDM$+\sum m_\nu$ free), respectively.
Across all dataset combinations considered, the presence of a free sterile neutrino mass is excluded at high statistical significance, typically exceeding the $5\sigma$ level.

\textit{These results indicate that the inclusion of a massive sterile neutrino is strongly disfavored by current cosmological data when compared to scenarios with the same number of free parameters in the absence of a sterile neutrino.}

\section{Conclusion}
\label{final}

In this work, we have presented a comprehensive observational study of the $w_\dagger$VCDM framework, focusing on its phenomenological viability and cosmological implications in the presence of extended neutrino sectors.  We have confronted the model with an extensive set of cosmological probes, including \textit{Planck} CMB measurements, DESI DR2 BAO data, and multiple SNIa compilations (PantheonPlus and Union3). Our analysis considered three increasingly general neutrino scenarios: a free total mass of the three active neutrinos $\sum m_\nu$ (\textbf{$w_{\dagger}$VCDM + $\sum m_\nu$}), a simultaneous variation of $\sum m_\nu$ and $N_{\rm eff}$ (\textbf{$w_{\dagger}$VCDM + $\sum m_\nu + N_{\rm eff}$}), and the presence of a fully thermalized sterile neutrino species (\textbf{$w_\dagger$VCDM + $m_{\rm sterile}$}). In all cases, we found that the extended $w_\dagger$VCDM framework provides an improved fit to the data compared to the $\Lambda$CDM paradigm, while remaining consistent with 
current observational constraints.

A central and robust result of our analysis is the strong preference for a negative transition amplitude, $\Delta<0$, across all considered dataset combinations and neutrino-sector extensions. This implies a transition from the phantom-like EoS ($w<-1$) at higher redshifts to a quintessence-like regime ($w>-1$) at late times. The transition redshift is tightly constrained to $z_\dagger \simeq 0.5$--$0.6$, indicating that current data favor a relatively late transition, occurring close to the onset of cosmic acceleration. \textit{(Could this be a new problem of cosmic coincidence?)} Remarkably, the inferred values of $\Delta$ and $z_\dagger$ remain highly stable under variations in the dataset combinations and neutrino-sector assumptions, and are fully consistent with previous analyses of this framework~\cite{Scherer:2025esj}.

Concerning the neutrino sector, the $w_\dagger$VCDM framework preserves the stringent bounds on the total mass of active neutrinos and remains consistent with the standard value of $N_{\rm eff}$ when allowed to vary. 

Within the sterile-neutrino extension, 
we obtain robust upper bounds on the effective sterile neutrino mass, $m_{\rm sterile} \lesssim 0.2$--$0.3~\mathrm{eV}$ (95\% CL), while leaving the DE transition parameters essentially unaffected. 
A key phenomenological outcome is the systematic increase in the inferred Hubble constant to $H_0 \simeq 71$--$72~\mathrm{km\,s^{-1}\,Mpc^{-1}}$, which significantly alleviates the Hubble tension, reducing it from $\sim5\sigma$ in $\Lambda$CDM to the $\sim2$--$2.5\sigma$ level, without invoking early DE or introducing instabilities.

We further observe  that model  comparison statistics robustly favor the $w_\dagger$VCDM framework over $\Lambda$CDM. All dataset combinations yield  $\Delta\chi^2_{\rm min} < 0$ and $\Delta\mathrm{AIC} <0$, demonstrating a statistically significant improvement in the fit even after accounting for the additional model parameters. 

In summary, the $w_\dagger$VCDM framework
emerges as a theoretically well-controlled extension of the $\Lambda$CDM paradigm. It provides a consistent description of current cosmological observations --- naturally accommodates late-time deviations from a cosmological constant --- remains fully compatible with constraints from neutrino physics --- and significantly alleviates the $H_0$ tension, while simultaneously offering 
an excellent fit to recent BAO and SNIa data.

We anticipate that future surveys with improved precision on the late-time expansion history and the growth of cosmic structures, together with independent probes of the relativistic particle content, will be crucial for testing the physical origin of the DE transition -- or, more generally, a possible transition within the dark sector -- and for assessing whether this phenomenological signature points toward new fundamental physics. It will also be of great interest to confront the model with additional independent datasets, particularly those sensitive to large-scale structure formation and nonlinear growth.


\begin{acknowledgments}
\noindent We thank the referee for the careful reading of the manuscript and for several constructive comments and suggestions, which helped us to improve the clarity and presentation of the manuscript. R.C.N. thanks the financial support from the Conselho Nacional de Desenvolvimento Científico e Tecnológico (CNPq, National Council for Scientific and Technological Development) under the project No. 304306/2022-3, and the Fundação de Amparo à Pesquisa do Estado do RS (FAPERGS, Research Support Foundation of the State of RS) for partial financial support under the project No. 23/2551-0000848-3. S.P acknowledges the partial support from the Department of Science and Technology (DST), Govt. of India under the Scheme   ``Fund for Improvement of S\&T Infrastructure (FIST)'' (File No. SR/FST/MS-I/2019/41). W.Y has been supported by the National Natural Science Foundation of China under Grant Nos. 12547110 and 12175096. 

\end{acknowledgments}
\bibliography{main}

@article{Bousis:2024rnb,
    author = "Bousis, Dimitrios and Perivolaropoulos, Leandros",
    title = "{Hubble tension tomography: BAO vs SN Ia distance tension}",
    eprint = "2405.07039",
    archivePrefix = "arXiv",
    primaryClass = "astro-ph.CO",
    doi = "10.1103/PhysRevD.110.103546",
    journal = "Phys. Rev. D",
    volume = "110",
    number = "10",
    pages = "103546",
    year = "2024"
}

@article{Pedrotti:2024kpn,
    author = "Pedrotti, Davide and Jiang, Jun-Qian and Escamilla, Luis A. and da Costa, Simony Santos and Vagnozzi, Sunny",
    title = "{Multidimensionality of the Hubble tension: The roles of {\ensuremath{\Omega}}m and {\ensuremath{\omega}}c}",
    eprint = "2408.04530",
    archivePrefix = "arXiv",
    primaryClass = "astro-ph.CO",
    doi = "10.1103/PhysRevD.111.023506",
    journal = "Phys. Rev. D",
    volume = "111",
    number = "2",
    pages = "023506",
    year = "2025"
}

@misc{DES:2025bxy,
    author = "Abbott, T. M. C. and others",
    collaboration = "DES",
    title = "{Dark Energy Survey: implications for cosmological expansion models from the final DES Baryon Acoustic Oscillation and Supernova data}",
    eprint = "2503.06712",
    archivePrefix = "arXiv",
    primaryClass = "astro-ph.CO",
    reportNumber = "FERMILAB-PUB-25-0127-PPD, DES-2024-0849, FERMILAB-PUB-25-0127-PPD",
    month = "3",
    year = "2025"
}

@misc{Ivanov:2026dvl,
    author = "Ivanov, Mikhail M. and Sullivan, James M. and Chen, Shi-Fan and Chudaykin, Anton and Maus, Mark and Philcox, Oliver H. E.",
    title = "{Reanalyzing DESI DR1: 4. Percent-Level Cosmological Constraints from Combined Probes and Robust Evidence for the Normal Neutrino Mass Hierarchy}",
    eprint = "2601.16165",
    archivePrefix = "arXiv",
    primaryClass = "astro-ph.CO",
    reportNumber = "MIT-CTP/5993",
    month = "1",
    year = "2026"
}

@article{Jhaveri:2025neg,
    author = "Jhaveri, Tanisha and Karwal, Tanvi and Hu, Wayne",
    title = "{Turning a negative neutrino mass into a positive optical depth}",
    eprint = "2504.21813",
    archivePrefix = "arXiv",
    primaryClass = "astro-ph.CO",
    doi = "10.1103/6vd2-rbfn",
    journal = "Phys. Rev. D",
    volume = "112",
    number = "4",
    pages = "043541",
    year = "2025"
}

@article{Lynch:2025ine,
    author = "Lynch, Gabriel P. and Knox, Lloyd",
    title = "{What{\textquoteright}s the matter with {\ensuremath{\Sigma}}m{\ensuremath{\nu}}?}",
    eprint = "2503.14470",
    archivePrefix = "arXiv",
    primaryClass = "astro-ph.CO",
    doi = "10.1103/613p-pph2",
    journal = "Phys. Rev. D",
    volume = "112",
    number = "8",
    pages = "083543",
    year = "2025"
}

@article{DESI:2025ffm,
    author = "Ahlen, S. P. and others",
    collaboration = "DESI",
    title = "{Positive Neutrino Masses with DESI DR2 via Matter Conversion to Dark Energy}",
    eprint = "2504.20338",
    archivePrefix = "arXiv",
    primaryClass = "astro-ph.CO",
    reportNumber = "FERMILAB-PUB-25-0288-PPD",
    doi = "10.1103/yb2k-kn7h",
    journal = "Phys. Rev. Lett.",
    volume = "135",
    number = "8",
    pages = "081003",
    year = "2025"
}

@article{Scherer:2025esj,
    author = "Scherer, Mateus and Sabogal, Miguel A. and Nunes, Rafael C. and De Felice, Antonio",
    title = "{Challenging the {\ensuremath{\Lambda}}CDM model: 5{\ensuremath{\sigma}} evidence for a dynamical dark energy late-time transition}",
    eprint = "2504.20664",
    archivePrefix = "arXiv",
    primaryClass = "astro-ph.CO",
    doi = "10.1103/n86r-sjgm",
    journal = "Phys. Rev. D",
    volume = "112",
    number = "4",
    pages = "043513",
    year = "2025"
}

@article{Green:2024xbb,
    author = "Green, Daniel and Meyers, Joel",
    title = "{Cosmological preference for a negative neutrino mass}",
    eprint = "2407.07878",
    archivePrefix = "arXiv",
    primaryClass = "astro-ph.CO",
    doi = "10.1103/PhysRevD.111.083507",
    journal = "Phys. Rev. D",
    volume = "111",
    number = "8",
    pages = "083507",
    year = "2025"
}

@article{Aoki:2018brq,
    author = "Aoki, Katsuki and De Felice, Antonio and Lin, Chunshan and Mukohyama, Shinji and Oliosi, Michele",
    title = "{Phenomenology in type-I minimally modified gravity}",
    eprint = "1810.01047",
    archivePrefix = "arXiv",
    primaryClass = "gr-qc",
    reportNumber = "WU-AP/1806/18, YITP-18-108, IPMU18-0158",
    doi = "10.1088/1475-7516/2019/01/017",
    journal = "JCAP",
    volume = "01",
    pages = "017",
    year = "2019"
}

@article{DeFelice:2015hla,
    author = "De Felice, Antonio and Mukohyama, Shinji",
    title = "{Minimal theory of massive gravity}",
    eprint = "1506.01594",
    archivePrefix = "arXiv",
    primaryClass = "hep-th",
    reportNumber = "YITP-15-48, IPMU15-0081",
    doi = "10.1016/j.physletb.2015.11.050",
    journal = "Phys. Lett. B",
    volume = "752",
    pages = "302--305",
    year = "2016"
}

@misc{Smith:2025uaq,
    author = "Smith, Adam and Mylova, Maria and van de Bruck, Carsten and Burgess, C. P. and Di Valentino, Eleonora",
    title = "{The Serendipitous Axiodilaton: A Self-Consistent Recombination-Era Solution to the Hubble Tension}",
    eprint = "2512.13544",
    archivePrefix = "arXiv",
    primaryClass = "astro-ph.CO",
    month = "12",
    year = "2025"
}

@misc{Sohail:2025mma,
    author = "Sohail, Sk. and Alam, Sonej and Hossain, Md. Wali",
    title = "{Observational constraints on early time non-phantom behaviour of dynamical dark energy}",
    eprint = "2512.19888",
    archivePrefix = "arXiv",
    primaryClass = "astro-ph.CO",
    month = "12",
    year = "2025"
}

@misc{Lee:2025axp,
    author = "Lee, Seokcheon",
    title = "{Controlled Tension Forecasting: Quantifying Cross-Probe Biases in $w_0w_a$CDM}",
    eprint = "2512.04130",
    archivePrefix = "arXiv",
    primaryClass = "astro-ph.CO",
    month = "12",
    year = "2025"
}

@misc{Efstratiou:2025iqi,
    author = "Efstratiou, Dimitrios and Paraskevas, Evangelos Achilleas and Perivolaropoulos, Leandros",
    title = "{Addressing the DESI DR2 Phantom-Crossing Anomaly and Enhanced $H_0$ Tension with Reconstructed Scalar-Tensor Gravity}",
    eprint = "2511.04610",
    archivePrefix = "arXiv",
    primaryClass = "astro-ph.CO",
    month = "11",
    year = "2025"
}

@misc{Smith:2025icl,
    author = {Smith, Adam and {\"O}z{\"u}lker, Emre and Di Valentino, Eleonora and van de Bruck, Carsten},
    title = "{Dynamical Dark Energy Meets Varying Electron Mass: Implications for Phantom Crossing and the Hubble Constant}",
    eprint = "2510.21931",
    archivePrefix = "arXiv",
    primaryClass = "astro-ph.CO",
    month = "10",
    year = "2025"
}

@misc{Liu:2025bss,
    author = "Liu, Rayne and Zhu, Yijie and Hu, Wayne and Miranda, Vivian",
    title = "{Phantom Mirage from Axion Dark Energy}",
    eprint = "2510.14957",
    archivePrefix = "arXiv",
    primaryClass = "astro-ph.CO",
    month = "10",
    year = "2025"
}

@misc{Li:2025muv,
    author = "Li, Tian-Nuo and Du, Guo-Hong and Li, Yun-He and Li, Yichao and Ling, Jia-Le and Zhang, Jing-Fei and Zhang, Xin",
    title = "{Updated constraints on interacting dark energy: A comprehensive analysis using multiple CMB probes, DESI DR2, and supernovae observations}",
    eprint = "2510.11363",
    archivePrefix = "arXiv",
    primaryClass = "astro-ph.CO",
    month = "10",
    year = "2025"
}

@article{Gomez-Valent:2025mfl,
    author = "G{\'o}mez-Valent, Adri{\`a} and Gonz{\'a}lez-Fuentes, Alex",
    title = "{Effective phantom divide crossing with standard and negative quintessence}",
    eprint = "2508.00621",
    archivePrefix = "arXiv",
    primaryClass = "astro-ph.CO",
    doi = "10.1016/j.physletb.2025.140096",
    journal = "Phys. Lett. B",
    volume = "872",
    pages = "140096",
    year = "2026"
}

@misc{Chaudhary:2025vzy,
    author = "Chaudhary, Himanshu and Capozziello, Salvatore and Sharma, Vipin Kumar and G{\'o}mez-Vargas, Isidro and Mustafa, G.",
    title = "{Evidence for Evolving Dark Energy from LRG1-2 and Low-$z$ SNe Ia Data}",
    eprint = "2508.10514",
    archivePrefix = "arXiv",
    primaryClass = "astro-ph.CO",
    month = "8",
    year = "2025"
}

@article{RoyChoudhury:2025iis,
    author = "Roy Choudhury, Shouvik and Okumura, Teppei and Umetsu, Keiichi",
    title = "{Cosmological Constraints on Nonphantom Dynamical Dark Energy with DESI Data Release 2 Baryon Acoustic Oscillations: A 3{\ensuremath{\sigma}}+ Lensing Anomaly}",
    eprint = "2509.26144",
    archivePrefix = "arXiv",
    primaryClass = "astro-ph.CO",
    doi = "10.3847/2041-8213/ae1a64",
    journal = "Astrophys. J. Lett.",
    volume = "994",
    number = "1",
    pages = "L26",
    year = "2025"
}

@misc{Chen:2025ywv,
    author = "Chen, Ruiqi and Cline, James M. and Muralidharan, Varun and Salewicz, Benjamin",
    title = "{Quintessential dark energy crossing the phantom divide}",
    eprint = "2508.19101",
    archivePrefix = "arXiv",
    primaryClass = "astro-ph.CO",
    month = "8",
    year = "2025"
}

@article{Fazzari:2025lzd,
    author = "Fazzari, Elisa and Giar{\`e}, William and Di Valentino, Eleonora",
    title = "{Cosmographic Footprints of Dynamical Dark Energy}",
    eprint = "2509.16196",
    archivePrefix = "arXiv",
    primaryClass = "astro-ph.CO",
    doi = "10.3847/2041-8213/ae2917",
    journal = "Astrophys. J. Lett.",
    volume = "996",
    number = "1",
    pages = "L5",
    year = "2026"
}

@misc{Montani:2025qnk,
    author = "Montani, Giovanni and Escamilla, Luis A. and Carlevaro, Nakia and Di Valentino, Eleonora",
    title = "{Decay of $f(R)$ quintessence into dark matter: mitigating the Hubble tension?}",
    eprint = "2512.20193",
    archivePrefix = "arXiv",
    primaryClass = "astro-ph.CO",
    month = "12",
    year = "2025"
}

@misc{deCruzPerez:2025dni,
    author = "de Cruz P{\'e}rez, Javier and G{\'o}mez-Valent, Adri{\`a} and Sol{\`a} Peracaula, Joan",
    title = "{Dynamical Dark Energy models in light of the latest observations}",
    eprint = "2512.20616",
    archivePrefix = "arXiv",
    primaryClass = "astro-ph.CO",
    month = "12",
    year = "2025"
}

@article{Pan:2023frx,
    author = "Pan, Supriya and Seto, Osamu and Takahashi, Tomo and Toda, Yo",
    title = "{Constraints on sterile neutrinos and the cosmological tensions}",
    eprint = "2312.15435",
    archivePrefix = "arXiv",
    primaryClass = "astro-ph.CO",
    reportNumber = "EPHOU-23-021",
    doi = "10.1103/PhysRevD.110.083524",
    journal = "Phys. Rev. D",
    volume = "110",
    number = "8",
    pages = "083524",
    year = "2024"
}

@misc{Chudaykin:2025lww,
    author = "Chudaykin, Anton and Ivanov, Mikhail M. and Philcox, Oliver H. E.",
    title = "{Reanalyzing DESI DR1: 2. Constraints on Dark Energy, Spatial Curvature, and Neutrino Masses}",
    eprint = "2511.20757",
    archivePrefix = "arXiv",
    primaryClass = "astro-ph.CO",
    reportNumber = "MIT-CTP/5960",
    month = "11",
    year = "2025"
}

@article{Silva:2025twg,
    author = "Silva, Emanuelly and Nunes, Rafael C.",
    title = "{Testing signatures of phantom crossing through full-shape galaxy clustering analysis}",
    eprint = "2507.13989",
    archivePrefix = "arXiv",
    primaryClass = "astro-ph.CO",
    doi = "10.1088/1475-7516/2025/11/078",
    journal = "JCAP",
    volume = "11",
    pages = "078",
    year = "2025"
}

@article{Sabogal:2025jbo,
    author = "Sabogal, Miguel A. and Nunes, Rafael C.",
    title = "{Robust evidence for dynamical dark energy from DESI galaxy-CMB lensing cross-correlation and geometric probes}",
    eprint = "2505.24465",
    archivePrefix = "arXiv",
    primaryClass = "astro-ph.CO",
    doi = "10.1088/1475-7516/2025/09/084",
    journal = "JCAP",
    volume = "09",
    pages = "084",
    year = "2025"
}

@misc{Reeves:2025xau,
    author = "Reeves, Alexander and Ferraro, Simone and Nicola, Andrina and Refregier, Alexandre",
    title = "{Multiprobe constraints on early and late time dark energy}",
    eprint = "2510.06114",
    archivePrefix = "arXiv",
    primaryClass = "astro-ph.CO",
    month = "10",
    year = "2025"
}

@misc{Ishak:2025cay,
    author = "Ishak, Mustapha and Medina-Varela, Leonel",
    title = "{Persistent and serious challenge to the $\Lambda$CDM throne: Evidence for dynamical dark energy rising from combinations of different types of datasets}",
    eprint = "2507.22856",
    archivePrefix = "arXiv",
    primaryClass = "astro-ph.CO",
    month = "7",
    year = "2025"
}

@misc{Ghedini:2025epp,
    author = "Ghedini, Pietro and Hajjar, Rasmi and Mena, Olga",
    title = "{Dark energy and neutrinos along the cosmic expansion history}",
    eprint = "2512.16781",
    archivePrefix = "arXiv",
    primaryClass = "astro-ph.CO",
    month = "12",
    year = "2025"
}

@misc{Barua:2025adv,
    author = "Barua, Shubham and Desai, Shantanu",
    title = "{Cosmological Constraints on Neutrino Masses in a Second-Order CPL Dark Energy Model}",
    eprint = "2508.16238",
    archivePrefix = "arXiv",
    primaryClass = "astro-ph.CO",
    month = "8",
    year = "2025"
}

@misc{SanchezLopez:2025uzw,
    author = "S{\'a}nchez L{\'o}pez, Samuel and Karam, Alexandros and Hazra, Dhiraj Kumar",
    title = "{Non-Minimally Coupled Quintessence in Light of DESI}",
    eprint = "2510.14941",
    archivePrefix = "arXiv",
    primaryClass = "astro-ph.CO",
    month = "10",
    year = "2025"
}

@misc{Artola:2025srt,
    author = "Artola, Mikel and Lazkoz, Ruth and Salzano, Vincenzo",
    title = "{A Spectrum of Cosmological Rips and Their Observational Signatures}",
    eprint = "2512.20383",
    archivePrefix = "arXiv",
    primaryClass = "astro-ph.CO",
    month = "12",
    year = "2025"
}

@article{Capozziello:2025qmh,
    author = "Capozziello, Salvatore and Chaudhary, Himanshu and Harko, Tiberiu and Mustafa, Ghulam",
    title = "{Is dark energy dynamical in the DESI era? A critical review}",
    eprint = "2512.10585",
    archivePrefix = "arXiv",
    primaryClass = "astro-ph.CO",
    doi = "10.1016/j.dark.2025.102196",
    journal = "Phys. Dark Univ.",
    volume = "51",
    pages = "102196",
    year = "2026"
}

@misc{Artola:2025zzb,
    author = "Artola, Mikel and Ayuso, Ismael and Lazkoz, Ruth and Salzano, Vincenzo",
    title = "{Is CPL dark energy a mirage?}",
    eprint = "2510.04191",
    archivePrefix = "arXiv",
    primaryClass = "astro-ph.CO",
    month = "10",
    year = "2025"
}

@misc{Wolf:2025acj,
    author = "Wolf, William J. and Ferreira, Pedro G. and Garc{\'\i}a-Garc{\'\i}a, Carlos",
    title = "{Cosmological constraints on Galileon dark energy with broken shift symmetry}",
    eprint = "2509.17586",
    archivePrefix = "arXiv",
    primaryClass = "astro-ph.CO",
    month = "9",
    year = "2025"
}

@misc{Adam:2025kve,
    author = "Adam, Husam and Hertzberg, Mark P. and Jim{\'e}nez-Aguilar, Daniel and Khan, Iman",
    title = "{Comparing Minimal and Non-Minimal Quintessence Models to 2025 DESI Data}",
    eprint = "2509.13302",
    archivePrefix = "arXiv",
    primaryClass = "astro-ph.CO",
    month = "9",
    year = "2025"
}

@article{Zhou:2025nkb,
    author = "Zhou, Sheng-Han and Li, Tian-Nuo and Du, Guo-Hong and Jiang, Jun-Qian and Zhang, Jing-Fei and Zhang, Xin",
    title = "{Measuring neutrino masses with joint JWST and DESI DR2 data}",
    eprint = "2509.10836",
    archivePrefix = "arXiv",
    primaryClass = "astro-ph.CO",
    doi = "10.1103/mtdg-hbqt",
    journal = "Phys. Rev. D",
    volume = "112",
    number = "12",
    pages = "123532",
    year = "2025"
}

@misc{Li:2025vuh,
    author = "Li, Tian-Nuo and Du, Guo-Hong and Zhou, Sheng-Han and Li, Yun-He and Zhang, Jing-Fei and Zhang, Xin",
    title = "{Robust evidence for dynamical dark energy in light of DESI DR2 and joint ACT, SPT, and Planck data}",
    eprint = "2511.22512",
    archivePrefix = "arXiv",
    primaryClass = "astro-ph.CO",
    month = "11",
    year = "2025"
}

@misc{Wu:2025vfs,
    author = "Wu, Peng-Ju and Li, Tian-Nuo and Du, Guo-Hong and Zhang, Xin",
    title = "{Observational challenges to holographic and Ricci dark energy paradigms: Insights from ACT DR6 and DESI DR2}",
    eprint = "2509.02945",
    archivePrefix = "arXiv",
    primaryClass = "astro-ph.CO",
    month = "9",
    year = "2025"
}

@article{diValentino:2022njd,
    author = "di Valentino, Eleonora and Gariazzo, Stefano and Mena, Olga",
    title = "{Model marginalized constraints on neutrino properties from cosmology}",
    eprint = "2207.05167",
    archivePrefix = "arXiv",
    primaryClass = "astro-ph.CO",
    doi = "10.1103/PhysRevD.106.043540",
    journal = "Phys. Rev. D",
    volume = "106",
    number = "4",
    pages = "043540",
    year = "2022"
}

@article{DiValentino:2021rjj,
    author = "Di Valentino, Eleonora and Gariazzo, Stefano and Giunti, Carlo and Mena, Olga and Pan, Supriya and Yang, Weiqiang",
    title = "{Minimal dark energy: Key to sterile neutrino and Hubble constant tensions?}",
    eprint = "2110.03990",
    archivePrefix = "arXiv",
    primaryClass = "astro-ph.CO",
    doi = "10.1103/PhysRevD.105.103511",
    journal = "Phys. Rev. D",
    volume = "105",
    number = "10",
    pages = "103511",
    year = "2022"
}

@article{Yang:2020ope,
    author = "Yang, Weiqiang and Di Valentino, Eleonora and Pan, Supriya and Mena, Olga",
    title = "{Emergent Dark Energy, neutrinos and cosmological tensions}",
    eprint = "2007.02927",
    archivePrefix = "arXiv",
    primaryClass = "astro-ph.CO",
    doi = "10.1016/j.dark.2020.100762",
    journal = "Phys. Dark Univ.",
    volume = "31",
    pages = "100762",
    year = "2021"
}

@article{Feleppa:2025clx,
    author = "Feleppa, Fabiano and Lambiase, Gaetano and Vagnozzi, Sunny",
    title = "{Imprints of screened dark energy on nonlocal quantum correlations}",
    eprint = "2508.18448",
    archivePrefix = "arXiv",
    primaryClass = "gr-qc",
    doi = "10.1103/y314-4x4s",
    journal = "Phys. Rev. D",
    volume = "112",
    number = "8",
    pages = "084011",
    year = "2025"
}

@article{Yao:2025wlx,
    author = "Yao, Zhibang and Ye, Gen and Silvestri, Alessandra",
    title = "{A general model for dark energy crossing the phantom divide}",
    eprint = "2508.01378",
    archivePrefix = "arXiv",
    primaryClass = "gr-qc",
    doi = "10.1088/1475-7516/2025/10/078",
    journal = "JCAP",
    volume = "10",
    pages = "078",
    year = "2025"
}

@article{Mishra:2025goj,
    author = "Mishra, Swagat S. and Matthewson, William L. and Sahni, Varun and Shafieloo, Arman and Shtanov, Yuri",
    title = "{Braneworld dark energy in light of DESI~DR2}",
    eprint = "2507.07193",
    archivePrefix = "arXiv",
    primaryClass = "astro-ph.CO",
    doi = "10.1088/1475-7516/2025/11/018",
    journal = "JCAP",
    volume = "11",
    pages = "018",
    year = "2025"
}

@article{Gialamas:2025pwv,
    author = {Gialamas, Ioannis D. and H{\"u}tsi, Gert and Raidal, Martti and Urrutia, Juan and Vasar, Martin and Veerm{\"a}e, Hardi},
    title = "{Quintessence and phantoms in light of DESI 2025}",
    eprint = "2506.21542",
    archivePrefix = "arXiv",
    primaryClass = "astro-ph.CO",
    doi = "10.1103/kdqc-y37v",
    journal = "Phys. Rev. D",
    volume = "112",
    number = "6",
    pages = "063551",
    year = "2025"
}

@misc{Ozulker:2025ehg,
    author = {{\"O}z{\"u}lker, Emre and Di Valentino, Eleonora and Giar{\`e}, William},
    title = "{Dark Energy Crosses the Line: Quantifying and Testing the Evidence for Phantom Crossing}",
    eprint = "2506.19053",
    archivePrefix = "arXiv",
    primaryClass = "astro-ph.CO",
    month = "6",
    year = "2025"
}

@article{Cline:2025sbt,
    author = "Cline, James M. and Muralidharan, Varun",
    title = "{Simple quintessence models in light of DESI-BAO observations}",
    eprint = "2506.13047",
    archivePrefix = "arXiv",
    primaryClass = "astro-ph.CO",
    doi = "10.1103/8z2m-nbv6",
    journal = "Phys. Rev. D",
    volume = "112",
    number = "6",
    pages = "063539",
    year = "2025"
}

@article{Mukherjee:2025ytj,
    author = "Mukherjee, Purba and Sen, Anjan A.",
    title = "{New expansion rate anomalies at characteristic redshifts geometrically determined using DESI-DR2 BAO and DES-SN5YR observations}",
    eprint = "2505.19083",
    archivePrefix = "arXiv",
    primaryClass = "astro-ph.CO",
    doi = "10.1088/1361-6633/ae082c",
    journal = "Rept. Prog. Phys.",
    volume = "88",
    number = "9",
    pages = "098401",
    year = "2025"
}

@article{Bayat:2025xfr,
    author = "Bayat, Zahra and Hertzberg, Mark P.",
    title = "{Examining quintessence models with DESI data}",
    eprint = "2505.18937",
    archivePrefix = "arXiv",
    primaryClass = "astro-ph.CO",
    doi = "10.1088/1475-7516/2025/08/065",
    journal = "JCAP",
    volume = "08",
    pages = "065",
    year = "2025"
}

@misc{Hussain:2025nqy,
    author = "Hussain, Saddam and Arora, Simran and Wang, Anzhong and Rose, Ben",
    title = "{Probing the Dynamics of Gaussian Dark Energy Equation of State Using DESI BAO}",
    eprint = "2505.09913",
    archivePrefix = "arXiv",
    primaryClass = "astro-ph.CO",
    doi = "10.1093/mnras/staf1924",
    month = "5",
    year = "2025"
}

@misc{Cheng:2025yue,
    author = "Cheng, Hanyu and Pan, Supriya and Di Valentino, Eleonora",
    title = "{Beyond Two Parameters: Revisiting Dark Energy with the Latest Cosmic Probes}",
    eprint = "2512.09866",
    archivePrefix = "arXiv",
    primaryClass = "astro-ph.CO",
    month = "12",
    year = "2025"
}

@article{Cheng:2025lod,
    author = "Cheng, Hanyu and Di Valentino, Eleonora and Escamilla, Luis A. and Sen, Anjan A. and Visinelli, Luca",
    title = "{Pressure parametrization of dark energy: first and second-order constraints with latest cosmological data}",
    eprint = "2505.02932",
    archivePrefix = "arXiv",
    primaryClass = "astro-ph.CO",
    reportNumber = "CA21106; CA21136",
    doi = "10.1088/1475-7516/2025/09/031",
    journal = "JCAP",
    volume = "09",
    pages = "031",
    year = "2025"
}

@misc{Toomey:2025yuy,
    author = "Toomey, Michael W. and Hughes, Ellie and Ivanov, Mikhail M. and Sullivan, James M.",
    title = "{Kinetic Mixing and the Phantom Illusion: Axion-Dilaton Quintessence in Light of DESI DR2}",
    eprint = "2511.23463",
    archivePrefix = "arXiv",
    primaryClass = "astro-ph.CO",
    reportNumber = "MIT-CTP/5963",
    month = "11",
    year = "2025"
}

@article{Paliathanasis:2025kmg,
    author = "Paliathanasis, Andronikos and Leon, Genly and Leyva, Yoelsy and Luciano, Giuseppe Gaetano and Abebe, Amare",
    title = "{Challenging {\ensuremath{\Lambda}}CDM with higher-order GUP corrections}",
    eprint = "2508.20644",
    archivePrefix = "arXiv",
    primaryClass = "gr-qc",
    doi = "10.1016/j.jheap.2025.100533",
    journal = "JHEAp",
    volume = "51",
    pages = "100533",
    year = "2026"
}

@article{Luciano:2025hjn,
    author = "Luciano, Giuseppe Gaetano and Paliathanasis, Andronikos and Saridakis, Emmanuel N.",
    title = "{Barrow and Tsallis entropies after the DESI DR2 BAO data}",
    eprint = "2504.12205",
    archivePrefix = "arXiv",
    primaryClass = "gr-qc",
    doi = "10.1088/1475-7516/2025/09/013",
    journal = "JCAP",
    volume = "09",
    pages = "013",
    year = "2025"
}

@misc{Du:2025xes,
    author = "Du, Guo-Hong and Li, Tian-Nuo and Wu, Peng-Ju and Zhang, Jing-Fei and Zhang, Xin",
    title = "{Cosmological Preference for a Positive Neutrino Mass at 2.7$\sigma$: A Joint Analysis of DESI DR2, DESY5, and DESY1 Data}",
    eprint = "2507.16589",
    archivePrefix = "arXiv",
    primaryClass = "astro-ph.CO",
    month = "7",
    year = "2025"
}

@article{Yang:2019uog,
    author = "Yang, Weiqiang and Pan, Supriya and Nunes, Rafael C. and Mota, David F.",
    title = "{Dark calling Dark: Interaction in the dark sector in presence of neutrino properties after Planck CMB final release}",
    eprint = "1910.08821",
    archivePrefix = "arXiv",
    primaryClass = "astro-ph.CO",
    doi = "10.1088/1475-7516/2020/04/008",
    journal = "JCAP",
    volume = "04",
    pages = "008",
    year = "2020"
}

@article{Yang:2017amu,
    author = "Yang, Weiqiang and Nunes, Rafael C. and Pan, Supriya and Mota, David F.",
    title = "{Effects of neutrino mass hierarchies on dynamical dark energy models}",
    eprint = "1703.02556",
    archivePrefix = "arXiv",
    primaryClass = "astro-ph.CO",
    doi = "10.1103/PhysRevD.95.103522",
    journal = "Phys. Rev. D",
    volume = "95",
    number = "10",
    pages = "103522",
    year = "2017"
}

@article{Wang:2025ker,
    author = "Wang, Deng and Mena, Olga and Di Valentino, Eleonora and Gariazzo, Stefano",
    title = "{Scale and redshift dependent limits on cosmic neutrino properties}",
    eprint = "2503.18745",
    archivePrefix = "arXiv",
    primaryClass = "astro-ph.CO",
    doi = "10.1103/6m5f-xn8r",
    journal = "Phys. Rev. D",
    volume = "112",
    number = "6",
    pages = "063555",
    year = "2025"
}

@article{Bertolez-Martinez:2024wez,
    author = "Bert{\'o}lez-Mart{\'\i}nez, Toni and Esteban, Ivan and Hajjar, Rasmi and Mena, Olga and Salvado, Jordi",
    title = "{Origin of cosmological neutrino mass bounds: background versus perturbations}",
    eprint = "2411.14524",
    archivePrefix = "arXiv",
    primaryClass = "astro-ph.CO",
    doi = "10.1088/1475-7516/2025/06/058",
    journal = "JCAP",
    volume = "06",
    pages = "058",
    year = "2025"
}

@article{Jiang:2024viw,
    author = "Jiang, Jun-Qian and Giar{\`e}, William and Gariazzo, Stefano and Dainotti, Maria Giovanna and Di Valentino, Eleonora and Mena, Olga and Pedrotti, Davide and da Costa, Simony Santos and Vagnozzi, Sunny",
    title = "{Neutrino cosmology after DESI: tightest mass upper limits, preference for the normal ordering, and tension with terrestrial observations}",
    eprint = "2407.18047",
    archivePrefix = "arXiv",
    primaryClass = "astro-ph.CO",
    doi = "10.1088/1475-7516/2025/01/153",
    journal = "JCAP",
    volume = "01",
    pages = "153",
    year = "2025"
}

@article{Shao:2025ohz,
    author = "Shao, Yue and Du, Guo-Hong and Li, Tian-Nuo and Zhang, Xin",
    title = "{Prospects for measuring neutrino mass with 21-cm forest}",
    eprint = "2501.00769",
    archivePrefix = "arXiv",
    primaryClass = "astro-ph.CO",
    doi = "10.1016/j.physletb.2025.139342",
    journal = "Phys. Lett. B",
    volume = "862",
    pages = "139342",
    year = "2025"
}

@article{Du:2024pai,
    author = "Du, Guo-Hong and Wu, Peng-Ju and Li, Tian-Nuo and Zhang, Xin",
    title = "{Impacts of dark energy on weighing neutrinos after DESI BAO}",
    eprint = "2407.15640",
    archivePrefix = "arXiv",
    primaryClass = "astro-ph.CO",
    doi = "10.1140/epjc/s10052-025-14094-0",
    journal = "Eur. Phys. J. C",
    volume = "85",
    number = "4",
    pages = "392",
    year = "2025"
}

@article{Liu:2020vgn,
    author = "Liu, Zhenjie and Miao, Haitao",
    title = "{Update constraints on neutrino mass and mass hierarchy in light of dark energy models}",
    eprint = "2002.05563",
    archivePrefix = "arXiv",
    primaryClass = "astro-ph.CO",
    reportNumber = "2050088",
    doi = "10.1142/S0218271820500881",
    journal = "Int. J. Mod. Phys. D",
    volume = "29",
    number = "13",
    pages = "2050088",
    year = "2020"
}

@article{CosmoVerseNetwork:2025alb,
    author = "Di Valentino, Eleonora and others",
    collaboration = "CosmoVerse Network",
    title = "{The CosmoVerse White Paper: Addressing observational tensions in cosmology with systematics and fundamental physics}",
    eprint = "2504.01669",
    archivePrefix = "arXiv",
    primaryClass = "astro-ph.CO",
    doi = "10.1016/j.dark.2025.101965",
    journal = "Phys. Dark Univ.",
    volume = "49",
    pages = "101965",
    year = "2025"
}

@misc{H0DN:2025lyy,
    author = "Casertano, Stefano and others",
    collaboration = "H0DN",
    title = "{The Local Distance Network: a community consensus report on the measurement of the Hubble constant at 1{\%} precision}",
    eprint = "2510.23823",
    archivePrefix = "arXiv",
    primaryClass = "astro-ph.CO",
    month = "10",
    year = "2025"
}

@article{Perivolaropoulos:2021jda,
    author = "Perivolaropoulos, Leandros and Skara, Foteini",
    title = "{Challenges for \ensuremath{\Lambda}CDM: An update}",
    eprint = "2105.05208",
    archivePrefix = "arXiv",
    primaryClass = "astro-ph.CO",
    doi = "10.1016/j.newar.2022.101659",
    journal = "New Astron. Rev.",
    volume = "95",
    pages = "101659",
    year = "2022"
}

@article{Arora:2025msq,
    author = "Arora, Simran and De Felice, Antonio and Mukohyama, Shinji",
    title = "{Dynamical dark energy parametrizations in VCDM}",
    eprint = "2508.03784",
    archivePrefix = "arXiv",
    primaryClass = "gr-qc",
    doi = "10.1103/l5bx-snl3",
    journal = "Phys. Rev. D",
    volume = "112",
    number = "12",
    pages = "123518",
    year = "2025"
}

@article{Blas:2011rf,
    author = "Blas, Diego and Lesgourgues, Julien and Tram, Thomas",
    title = "{The Cosmic Linear Anisotropy Solving System (CLASS) II: Approximation schemes}",
    eprint = "1104.2933",
    archivePrefix = "arXiv",
    primaryClass = "astro-ph.CO",
    reportNumber = "CERN-PH-TH-2011-082, LAPTH-010-11",
    doi = "10.1088/1475-7516/2011/07/034",
    journal = "JCAP",
    volume = "07",
    pages = "034",
    year = "2011"
}

@ARTICLE{1100705,
  author={Akaike, H.},
  journal={IEEE Transactions on Automatic Control}, 
  title={A new look at the statistical model identification}, 
  year={1974},
  volume={19},
  number={6},
  pages={716-723},
  doi={10.1109/TAC.1974.1100705}}

@article{Planck:2013pxb,
    author = "Ade, P. A. R. and others",
    collaboration = "Planck",
    title = "{Planck 2013 results. XVI. Cosmological parameters}",
    eprint = "1303.5076",
    archivePrefix = "arXiv",
    primaryClass = "astro-ph.CO",
    reportNumber = "CERN-PH-TH-2013-129",
    doi = "10.1051/0004-6361/201321591",
    journal = "Astron. Astrophys.",
    volume = "571",
    pages = "A16",
    year = "2014"
}

@article{Gelman_1992,
    author = "Gelman, Andrew and Rubin, Donald B.",
    title = "{Inference from Iterative Simulation Using Multiple Sequences}",
    doi = "10.1214/ss/1177011136",
    journal = "Statist. Sci.",
    volume = "7",
    pages = "457--472",
    year = "1992"
}

@article{Planck:2018lbu,
    author = "Aghanim, N. and others",
    collaboration = "Planck",
    title = "{Planck 2018 results. VIII. Gravitational lensing}",
    eprint = "1807.06210",
    archivePrefix = "arXiv",
    primaryClass = "astro-ph.CO",
    doi = "10.1051/0004-6361/201833886",
    journal = "Astron. Astrophys.",
    volume = "641",
    pages = "A8",
    year = "2020"
}

@article{DESI:2024mwx,
    author = "Adame, A. G. and others",
    collaboration = "DESI",
    title = "{DESI 2024 VI: cosmological constraints from the measurements of baryon acoustic oscillations}",
    eprint = "2404.03002",
    archivePrefix = "arXiv",
    primaryClass = "astro-ph.CO",
    reportNumber = "FERMILAB-PUB-24-0154-PPD",
    doi = "10.1088/1475-7516/2025/02/021",
    journal = "JCAP",
    volume = "02",
    pages = "021",
    year = "2025"
}

@article{Addison:2015wyg,
    author = "Addison, G. E. and Huang, Y. and Watts, D. J. and Bennett, C. L. and Halpern, M. and Hinshaw, G. and Weiland, J. L.",
    title = "{Quantifying discordance in the 2015 Planck CMB spectrum}",
    eprint = "1511.00055",
    archivePrefix = "arXiv",
    primaryClass = "astro-ph.CO",
    doi = "10.3847/0004-637X/818/2/132",
    journal = "Astrophys. J.",
    volume = "818",
    number = "2",
    pages = "132",
    year = "2016"
}

@misc{Akarsu:2024qsi,
    author = "Akarsu, Ozgur and De Felice, Antonio and Di Valentino, Eleonora and Kumar, Suresh and Nunes, Rafael C. and Ozulker, Emre and Vazquez, J. Alberto and Yadav, Anita",
    title = "{$\Lambda_{\rm s}$CDM cosmology from a type-II minimally modified gravity}",
    eprint = "2402.07716",
    archivePrefix = "arXiv",
    primaryClass = "astro-ph.CO",
    reportNumber = "YITP-24-18",
    month = "2",
    year = "2024"
}

@article{Nunes:2021ipq,
    author = "Nunes, Rafael C. and Vagnozzi, Sunny",
    title = "{Arbitrating the S8 discrepancy with growth rate measurements from redshift-space distortions}",
    eprint = "2106.01208",
    archivePrefix = "arXiv",
    primaryClass = "astro-ph.CO",
    doi = "10.1093/mnras/stab1613",
    journal = "Mon. Not. Roy. Astron. Soc.",
    volume = "505",
    number = "4",
    pages = "5427--5437",
    year = "2021"
}

@misc{Dvorkin:2019jgs,
    author = "Dvorkin, Cora and others",
    title = "{Neutrino Mass from Cosmology: Probing Physics Beyond the Standard Model}",
    eprint = "1903.03689",
    archivePrefix = "arXiv",
    primaryClass = "astro-ph.CO",
    month = "3",
    year = "2019"
}

@article{Gariazzo:2024sil,
    author = "Gariazzo, Stefano and Giar{\`e}, William and Mena, Olga and Di Valentino, Eleonora",
    title = "{How robust are the parameter constraints extending the {\ensuremath{\Lambda}}CDM model?}",
    eprint = "2404.11182",
    archivePrefix = "arXiv",
    primaryClass = "astro-ph.CO",
    doi = "10.1103/PhysRevD.111.023540",
    journal = "Phys. Rev. D",
    volume = "111",
    number = "2",
    pages = "023540",
    year = "2025"
}

@article{Wang:2024hen,
    author = "Wang, Deng and Mena, Olga and Di Valentino, Eleonora and Gariazzo, Stefano",
    title = "{Updating neutrino mass constraints with background measurements}",
    eprint = "2405.03368",
    archivePrefix = "arXiv",
    primaryClass = "astro-ph.CO",
    doi = "10.1103/PhysRevD.110.103536",
    journal = "Phys. Rev. D",
    volume = "110",
    number = "10",
    pages = "103536",
    year = "2024"
}

@article{Kamionkowski:2022pkx,
    author = "Kamionkowski, Marc and Riess, Adam G.",
    title = "{The Hubble Tension and Early Dark Energy}",
    eprint = "2211.04492",
    archivePrefix = "arXiv",
    primaryClass = "astro-ph.CO",
    doi = "10.1146/annurev-nucl-111422-024107",
    journal = "Ann. Rev. Nucl. Part. Sci.",
    volume = "73",
    pages = "153--180",
    year = "2023"
}

@article{DiValentino:2021izs,
    author = "Di Valentino, Eleonora and Mena, Olga and Pan, Supriya and Visinelli, Luca and Yang, Weiqiang and Melchiorri, Alessandro and Mota, David F. and Riess, Adam G. and Silk, Joseph",
    title = "{In the realm of the Hubble tension\textemdash{}a review of solutions}",
    eprint = "2103.01183",
    archivePrefix = "arXiv",
    primaryClass = "astro-ph.CO",
    reportNumber = "IPPP/20/108",
    doi = "10.1088/1361-6382/ac086d",
    journal = "Class. Quant. Grav.",
    volume = "38",
    number = "15",
    pages = "153001",
    year = "2021"
}

@article{Planck:2018vyg,
    author = "Aghanim, N. and others",
    collaboration = "Planck",
    title = "{Planck 2018 results. VI. Cosmological parameters}",
    eprint = "1807.06209",
    archivePrefix = "arXiv",
    primaryClass = "astro-ph.CO",
    doi = "10.1051/0004-6361/201833910",
    journal = "Astron. Astrophys.",
    volume = "641",
    pages = "A6",
    year = "2020",
    note = "[Erratum: Astron.Astrophys. 652, C4 (2021)]"
}

@article{Planck:2019nip,
    author = "Aghanim, N. and others",
    collaboration = "Planck",
    title = "{Planck 2018 results. V. CMB power spectra and likelihoods}",
    eprint = "1907.12875",
    archivePrefix = "arXiv",
    primaryClass = "astro-ph.CO",
    doi = "10.1051/0004-6361/201936386",
    journal = "Astron. Astrophys.",
    volume = "641",
    pages = "A5",
    year = "2020"
}

@article{ACT:2020gnv,
    author = "Aiola, Simone and others",
    collaboration = "ACT",
    title = "{The Atacama Cosmology Telescope: DR4 Maps and Cosmological Parameters}",
    eprint = "2007.07288",
    archivePrefix = "arXiv",
    primaryClass = "astro-ph.CO",
    doi = "10.1088/1475-7516/2020/12/047",
    journal = "JCAP",
    volume = "12",
    pages = "047",
    year = "2020"
}

@article{Riess:2021jrx,
    author = "Riess, Adam G. and others",
    title = "{A Comprehensive Measurement of the Local Value of the Hubble Constant with 1 km s$^{−1}$ Mpc$^{−1}$ Uncertainty from the Hubble Space Telescope and the SH0ES Team}",
    eprint = "2112.04510",
    archivePrefix = "arXiv",
    primaryClass = "astro-ph.CO",
    doi = "10.3847/2041-8213/ac5c5b",
    journal = "Astrophys. J. Lett.",
    volume = "934",
    number = "1",
    pages = "L7",
    year = "2022"
}

@article{DESI:2025zgx,
    author = "Abdul Karim, M. and others",
    collaboration = "DESI",
    title = "{DESI DR2 results. II. Measurements of baryon acoustic oscillations and cosmological constraints}",
    eprint = "2503.14738",
    archivePrefix = "arXiv",
    primaryClass = "astro-ph.CO",
    reportNumber = "FERMILAB-PUB-25-0169-PPD",
    doi = "10.1103/tr6y-kpc6",
    journal = "Phys. Rev. D",
    volume = "112",
    number = "8",
    pages = "083515",
    year = "2025"
}

@article{DESI:2025zpo,
    author = "Abdul Karim, M. and others",
    collaboration = "DESI",
    title = "{DESI DR2 results. I. Baryon acoustic oscillations from the Lyman alpha forest}",
    eprint = "2503.14739",
    archivePrefix = "arXiv",
    primaryClass = "astro-ph.CO",
    reportNumber = "FERMILAB-PUB-25-0167-PPD",
    doi = "10.1103/2wwn-xjm5",
    journal = "Phys. Rev. D",
    volume = "112",
    number = "8",
    pages = "083514",
    year = "2025"
}

@article{Escudero:2024uea,
    author = "Escudero, Helena Garc\'\i{}a and Abazajian, Kevork N.",
    title = "{Status of neutrino cosmology: Standard \ensuremath{\Lambda}CDM, extensions, and tensions}",
    eprint = "2412.05451",
    archivePrefix = "arXiv",
    primaryClass = "astro-ph.CO",
    reportNumber = "UCI-HEP-TR-2024-20",
    doi = "10.1103/PhysRevD.111.043520",
    journal = "Phys. Rev. D",
    volume = "111",
    number = "4",
    pages = "043520",
    year = "2025"
}

@article{Feng:2025mlo,
    author = "Feng, Lu and Li, Tian-Nuo and Du, Guo-Hong and Zhang, Jing-Fei and Zhang, Xin",
    title = "{A search for sterile neutrinos in interacting dark energy models using DESI baryon acoustic oscillations and DES supernovae data}",
    eprint = "2503.10423",
    archivePrefix = "arXiv",
    primaryClass = "astro-ph.CO",
    doi = "10.1016/j.dark.2025.101935",
    journal = "Phys. Dark Univ.",
    volume = "48",
    pages = "101935",
    year = "2025"
}

@article{RoyChoudhury:2025dhe,
    author = "Roy Choudhury, Shouvik",
    title = "{Cosmology in Extended Parameter Space with DESI Data Release 2 Baryon Acoustic Oscillations: A 2{\ensuremath{\sigma}}+ Detection of Nonzero Neutrino Masses with an Update on Dynamical Dark Energy and Lensing Anomaly}",
    eprint = "2504.15340",
    archivePrefix = "arXiv",
    primaryClass = "astro-ph.CO",
    doi = "10.3847/2041-8213/ade1cc",
    journal = "Astrophys. J. Lett.",
    volume = "986",
    number = "2",
    pages = "L31",
    year = "2025"
}

@article{RoyChoudhury:2024wri,
    author = "Roy Choudhury, Shouvik and Okumura, Teppei",
    title = "{Updated Cosmological Constraints in Extended Parameter Space with Planck PR4, DESI Baryon Acoustic Oscillations, and Supernovae: Dynamical Dark Energy, Neutrino Masses, Lensing Anomaly, and the Hubble Tension}",
    eprint = "2409.13022",
    archivePrefix = "arXiv",
    primaryClass = "astro-ph.CO",
    doi = "10.3847/2041-8213/ad8c26",
    journal = "Astrophys. J. Lett.",
    volume = "976",
    number = "1",
    pages = "L11",
    year = "2024"
}

@article{Elbers:2025vlz,
    author = "Elbers, W. and others",
    title = "{Constraints on neutrino physics from DESI DR2 BAO and DR1 full shape}",
    eprint = "2503.14744",
    archivePrefix = "arXiv",
    primaryClass = "astro-ph.CO",
    reportNumber = "FERMILAB-PUB-25-0168-PPD",
    doi = "10.1103/w9pk-xsk7",
    journal = "Phys. Rev. D",
    volume = "112",
    number = "8",
    pages = "083513",
    year = "2025"
}

@article{Brout:2022vxf,
    author = "Brout, Dillon and others",
    title = "{The Pantheon+ Analysis: Cosmological Constraints}",
    eprint = "2202.04077",
    archivePrefix = "arXiv",
    primaryClass = "astro-ph.CO",
    doi = "10.3847/1538-4357/ac8e04",
    journal = "Astrophys. J.",
    volume = "938",
    number = "2",
    pages = "110",
    year = "2022"
}

@article{Rubin:2023jdq,
    author = "Rubin, David and others",
    title = "{Union Through UNITY: Cosmology with 2,000 SNe Using a Unified Bayesian Framework}",
    eprint = "2311.12098",
    archivePrefix = "arXiv",
    primaryClass = "astro-ph.CO",
    doi = "10.3847/1538-4357/adc0a5",
    journal = "Astrophys. J.",
    volume = "986",
    number = "2",
    pages = "231",
    year = "2025"
}

@misc{DiValentino:2024xsv,
    author = "Di Valentino, Eleonora and Gariazzo, Stefano and Mena, Olga",
    title = "{Neutrinos in Cosmology}",
    eprint = "2404.19322",
    archivePrefix = "arXiv",
    primaryClass = "astro-ph.CO",
    month = "4",
    year = "2024"
}

@misc{Vagnozzi:2019utt,
    author = "Vagnozzi, Sunny",
    title = "{Cosmological searches for the neutrino mass scale and mass ordering}",
    eprint = "1907.08010",
    archivePrefix = "arXiv",
    primaryClass = "astro-ph.CO",
    month = "7",
    year = "2019"
}

@article{Lesgourgues:2014zoa,
    author = "Lesgourgues, Julien and Pastor, Sergio",
    title = "{Neutrino cosmology and Planck}",
    eprint = "1404.1740",
    archivePrefix = "arXiv",
    primaryClass = "hep-ph",
    reportNumber = "MITP-14-015",
    doi = "10.1088/1367-2630/16/6/065002",
    journal = "New J. Phys.",
    volume = "16",
    pages = "065002",
    year = "2014"
}

@article{Gariazzo:2022ahe,
    author = "Gariazzo, Stefano and others",
    title = "{Neutrino mass and mass ordering: no conclusive evidence for normal ordering}",
    eprint = "2205.02195",
    archivePrefix = "arXiv",
    primaryClass = "hep-ph",
    doi = "10.1088/1475-7516/2022/10/010",
    journal = "JCAP",
    volume = "10",
    pages = "010",
    year = "2022"
}

@article{Naredo-Tuero:2024sgf,
    author = "Naredo-Tuero, Daniel and Escudero, Miguel and Fern{\'a}ndez-Mart{\'\i}nez, Enrique and Marcano, Xabier and Poulin, Vivian",
    title = "{Critical look at the cosmological neutrino mass bound}",
    eprint = "2407.13831",
    archivePrefix = "arXiv",
    primaryClass = "astro-ph.CO",
    reportNumber = "CERN-TH-2024-115, IFT-UAM/CSIC-24-106",
    doi = "10.1103/PhysRevD.110.123537",
    journal = "Phys. Rev. D",
    volume = "110",
    number = "12",
    pages = "123537",
    year = "2024"
}

@article{Elbers:2024sha,
    author = "Elbers, Willem and Frenk, Carlos S. and Jenkins, Adrian and Li, Baojiu and Pascoli, Silvia",
    title = "{Negative neutrino masses as a mirage of dark energy}",
    eprint = "2407.10965",
    archivePrefix = "arXiv",
    primaryClass = "astro-ph.CO",
    doi = "10.1103/PhysRevD.111.063534",
    journal = "Phys. Rev. D",
    volume = "111",
    number = "6",
    pages = "063534",
    year = "2025"
}

@article{TopicalConvenersKNAbazajianJECarlstromATLee:2013bxd,
    author = "Abazajian, K. N. and others",
    collaboration = "Topical Conveners: K.N. Abazajian, J.E. Carlstrom, A.T. Lee",
    title = "{Neutrino Physics from the Cosmic Microwave Background and Large Scale Structure}",
    eprint = "1309.5383",
    archivePrefix = "arXiv",
    primaryClass = "astro-ph.CO",
    reportNumber = "FERMILAB-PUB-13-438-A",
    doi = "10.1016/j.astropartphys.2014.05.014",
    journal = "Astropart. Phys.",
    volume = "63",
    pages = "66--80",
    year = "2015"
}

@article{Lesgourgues:2006nd,
    author = "Lesgourgues, Julien and Pastor, Sergio",
    title = "{Massive neutrinos and cosmology}",
    eprint = "astro-ph/0603494",
    archivePrefix = "arXiv",
    reportNumber = "LAPTH-1131-05, IFIC-05-59",
    doi = "10.1016/j.physrep.2006.04.001",
    journal = "Phys. Rept.",
    volume = "429",
    pages = "307--379",
    year = "2006"
}

@article{DeFelice:2020eju,
    author = "De Felice, Antonio and Doll, Andreas and Mukohyama, Shinji",
    title = "{A theory of type-II minimally modified gravity}",
    eprint = "2004.12549",
    archivePrefix = "arXiv",
    primaryClass = "gr-qc",
    reportNumber = "YITP-20-55, IPMU20-0040",
    doi = "10.1088/1475-7516/2020/09/034",
    journal = "JCAP",
    volume = "09",
    pages = "034",
    year = "2020"
}

@article{DeFelice:2020cpt,
    author = "De Felice, Antonio and Mukohyama, Shinji and Pookkillath, Masroor C.",
    title = "{Addressing $H_0$ tension by means of VCDM}",
    eprint = "2009.08718",
    archivePrefix = "arXiv",
    primaryClass = "astro-ph.CO",
    reportNumber = "YITP-20-117, IPMU 20-0098",
    doi = "10.1016/j.physletb.2021.136201",
    journal = "Phys. Lett. B",
    volume = "816",
    pages = "136201",
    year = "2021",
    note = "[Erratum: Phys.Lett.B 818, 136364 (2021)]"
}

@article{DeFelice:2021xps,
    author = "De Felice, Antonio and Mukohyama, Shinji and Pookkillath, Masroor C.",
    title = "{Static, spherically symmetric objects in type-II minimally modified gravity}",
    eprint = "2110.14496",
    archivePrefix = "arXiv",
    primaryClass = "gr-qc",
    reportNumber = "YITP-21-127, IPMU21-0068",
    doi = "10.1103/PhysRevD.105.104013",
    journal = "Phys. Rev. D",
    volume = "105",
    number = "10",
    pages = "104013",
    year = "2022"
}

@article{DeFelice:2022riv,
    author = "De Felice, Antonio and Maeda, Kei-ichi and Mukohyama, Shinji and Pookkillath, Masroor C.",
    title = "{Gravitational collapse and formation of a black hole in a type II minimally modified gravity theory}",
    eprint = "2211.14760",
    archivePrefix = "arXiv",
    primaryClass = "gr-qc",
    reportNumber = "YITP-22-142, IPMU22-0062",
    doi = "10.1088/1475-7516/2023/03/030",
    journal = "JCAP",
    volume = "03",
    pages = "030",
    year = "2023"
}

@article{Jalali:2023wqh,
    author = "Jalali, Atabak Fathe and Martens, Paul and Mukohyama, Shinji",
    title = "{Spherical scalar collapse in a type-II minimally modified gravity}",
    eprint = "2306.10672",
    archivePrefix = "arXiv",
    primaryClass = "gr-qc",
    reportNumber = "YITP-23-75, IPMU23-0023",
    doi = "10.1103/PhysRevD.109.044053",
    journal = "Phys. Rev. D",
    volume = "109",
    number = "4",
    pages = "044053",
    year = "2024"
}

@article{DeFelice:2022uxv,
    author = "De Felice, Antonio and Maeda, Kei-ichi and Mukohyama, Shinji and Pookkillath, Masroor C.",
    title = "{Comparison of two theories of Type-IIa minimally modified gravity}",
    eprint = "2204.08294",
    archivePrefix = "arXiv",
    primaryClass = "gr-qc",
    reportNumber = "YITP-22-39, IPMU22-0020",
    doi = "10.1103/PhysRevD.106.024028",
    journal = "Phys. Rev. D",
    volume = "106",
    number = "2",
    pages = "024028",
    year = "2022"
}

@article{Torrado:2020dgo,
    author = "Torrado, Jesus and Lewis, Antony",
    title = "{Cobaya: Code for Bayesian Analysis of hierarchical physical models}",
    eprint = "2005.05290",
    archivePrefix = "arXiv",
    primaryClass = "astro-ph.IM",
    reportNumber = "TTK-20-15",
    doi = "10.1088/1475-7516/2021/05/057",
    journal = "JCAP",
    volume = "05",
    pages = "057",
    year = "2021"
}

@article{Akarsu:2024eoo,
    author = {Akarsu, \"Ozg\"ur and De Felice, Antonio and Di Valentino, Eleonora and Kumar, Suresh and Nunes, Rafael C. and \"Oz\"ulker, Emre and Vazquez, J. Alberto and Yadav, Anita},
    title = "{Cosmological constraints on \ensuremath{\Lambda}sCDM scenario in a type II minimally modified gravity}",
    eprint = "2406.07526",
    archivePrefix = "arXiv",
    primaryClass = "astro-ph.CO",
    reportNumber = "YITP-24-57",
    doi = "10.1103/PhysRevD.110.103527",
    journal = "Phys. Rev. D",
    volume = "110",
    number = "10",
    pages = "103527",
    year = "2024"
}

@misc{Nair:2025uyn,
    author = "Nair, Gowri S. and Chakraborty, Amlan and Amendola, Luca and Das, Subinoy",
    title = "{Neutrino mass constraints in the context of 4-parameter dark energy equation of state and DESI DR2 observations}",
    eprint = "2512.08752",
    archivePrefix = "arXiv",
    primaryClass = "astro-ph.CO",
    month = "12",
    year = "2025"
}

@article{Giare:2025ath,
    author = "Giar{\`e}, William and Mena, Olga and Specogna, Enrico and Di Valentino, Eleonora",
    title = "{Neutrino mass tension or suppressed growth rate of matter perturbations?}",
    eprint = "2507.01848",
    archivePrefix = "arXiv",
    primaryClass = "astro-ph.CO",
    doi = "10.1103/njfc-pd1w",
    journal = "Phys. Rev. D",
    volume = "112",
    number = "10",
    pages = "103520",
    year = "2025"
}

@misc{Hussain:2025vbo,
    author = "Hussain, Saddam and Arora, Simran and Rana, Yamuna and Rose, Benjamin and Wang, Anzhong",
    title = "{Interacting Scalar Fields as Dark Energy and Dark Matter in Einstein scalar Gauss Bonnet Gravity}",
    eprint = "2507.05207",
    archivePrefix = "arXiv",
    primaryClass = "gr-qc",
    month = "7",
    year = "2025"
}

@article{Abazajian:2017tcc,
    author = "Abazajian, Kevork N.",
    title = "{Sterile neutrinos in cosmology}",
    eprint = "1705.01837",
    archivePrefix = "arXiv",
    primaryClass = "hep-ph",
    reportNumber = "UCI-TR-2017-03",
    doi = "10.1016/j.physrep.2017.10.003",
    journal = "Phys. Rept.",
    volume = "711-712",
    pages = "1--28",
    year = "2017"
}

@article{Hu:2025lrl,
    author = "Hu, Rui and Chu, Ming-chung and Yeung, Shek and Zhang, Wangzheng",
    title = "{Impact of light sterile neutrinos on cosmological large scale structure}",
    eprint = "2501.16908",
    archivePrefix = "arXiv",
    primaryClass = "astro-ph.CO",
    doi = "10.1088/1475-7516/2025/06/014",
    journal = "JCAP",
    volume = "06",
    pages = "014",
    year = "2025"
}

@article{Boyarsky:2009ix,
    author = "Boyarsky, Alexey and Ruchayskiy, Oleg and Shaposhnikov, Mikhail",
    title = "{The Role of sterile neutrinos in cosmology and astrophysics}",
    eprint = "0901.0011",
    archivePrefix = "arXiv",
    primaryClass = "hep-ph",
    doi = "10.1146/annurev.nucl.010909.083654",
    journal = "Ann. Rev. Nucl. Part. Sci.",
    volume = "59",
    pages = "191--214",
    year = "2009"
}

@article{Feng:2021ipq,
    author = "Feng, Lu and Guo, Rui-Yun and Zhang, Jing-Fei and Zhang, Xin",
    title = "{Cosmological search for sterile neutrinos after Planck 2018}",
    eprint = "2109.06111",
    archivePrefix = "arXiv",
    primaryClass = "astro-ph.CO",
    doi = "10.1016/j.physletb.2022.136940",
    journal = "Phys. Lett. B",
    volume = "827",
    pages = "136940",
    year = "2022"
}

@misc{Du:2025iow,
    author = "Du, Guo-Hong and Li, Tian-Nuo and Wu, Peng-Ju and Feng, Lu and Zhou, Sheng-Han and Zhang, Jing-Fei and Zhang, Xin",
    title = "{Cosmological search for sterile neutrinos after DESI 2024}",
    eprint = "2501.10785",
    archivePrefix = "arXiv",
    primaryClass = "astro-ph.CO",
    month = "1",
    year = "2025"
}

@article{Feng:2025wbz,
    author = "Feng, Lu and Han, Tao and Zhang, Jing-Fei and Zhang, Xin",
    title = "{Prospects for searching for sterile neutrinos in dynamical dark energy cosmologies using joint observations of gravitational waves and {\ensuremath{\gamma}}-ray bursts*}",
    eprint = "2507.17315",
    archivePrefix = "arXiv",
    primaryClass = "astro-ph.CO",
    doi = "10.1088/1674-1137/ae0b43",
    journal = "Chin. Phys.",
    volume = "50",
    number = "1",
    pages = "015105",
    year = "2026"
}

@article{Benso:2024qrg,
    author = "Benso, Cristina and Schwetz, Thomas and Vatsyayan, Drona",
    title = "{Large neutrino mass in cosmology and keV sterile neutrino dark matter from a dark sector}",
    eprint = "2410.23926",
    archivePrefix = "arXiv",
    primaryClass = "hep-ph",
    doi = "10.1088/1475-7516/2025/04/054",
    journal = "JCAP",
    volume = "04",
    pages = "054",
    year = "2025"
}

@misc{GarciaEscudero:2025orc,
    author = "Garc{\'\i}a Escudero, Helena and Abazajian, Kevork N.",
    title = "{Extra Radiation Cosmologies: Implications of the Hubble Tension for eV-scale Neutrinos}",
    eprint = "2509.25478",
    archivePrefix = "arXiv",
    primaryClass = "hep-ph",
    reportNumber = "UCI-HEP-TR-2025-21",
    month = "9",
    year = "2025"
}

@article{Escudero:2022rbq,
    author = "Escudero, Helena Garc{\'\i}a and Kuo, Jui-Lin and Keeley, Ryan E. and Abazajian, Kevork N.",
    title = "{Early or phantom dark energy, self-interacting, extra, or massive neutrinos, primordial magnetic fields, or a curved universe: An exploration of possible solutions to the H0 and {\ensuremath{\sigma}}8 problems}",
    eprint = "2208.14435",
    archivePrefix = "arXiv",
    primaryClass = "astro-ph.CO",
    reportNumber = "UCI-HEP-TR-2022-06",
    doi = "10.1103/PhysRevD.106.103517",
    journal = "Phys. Rev. D",
    volume = "106",
    number = "10",
    pages = "103517",
    year = "2022"
}

@article{Hagstotz:2020ukm,
    author = "Hagstotz, Steffen and de Salas, Pablo F. and Gariazzo, Stefano and Gerbino, Martina and Lattanzi, Massimiliano and Vagnozzi, Sunny and Freese, Katherine and Pastor, Sergio",
    title = "{Bounds on light sterile neutrino mass and mixing from cosmology and laboratory searches}",
    eprint = "2003.02289",
    archivePrefix = "arXiv",
    primaryClass = "astro-ph.CO",
    doi = "10.1103/PhysRevD.104.123524",
    journal = "Phys. Rev. D",
    volume = "104",
    number = "12",
    pages = "123524",
    year = "2021"
}

@article{Dhuria:2023yrw,
    author = "Dhuria, Mansi and Pradhan, Abinas",
    title = "{Synergy between Hubble tension motivated self-interacting neutrinos and KeV-sterile neutrino dark matter}",
    eprint = "2301.09552",
    archivePrefix = "arXiv",
    primaryClass = "hep-ph",
    doi = "10.1103/PhysRevD.107.123030",
    journal = "Phys. Rev. D",
    volume = "107",
    number = "12",
    pages = "123030",
    year = "2023"
}

@article{KATRIN:2025Science,
  author        = {{KATRIN Collaboration} and Aker, M. and others},
  title         = {Direct neutrino mass constraints from tritium $\beta$-decay kinematics},
  journal       = {Science},
  volume        = {389},
  pages         = {64--69},
  year          = {2025}
}

@article{RaveriHu:2018,
  author        = {Raveri, Marco and Hu, Wayne},
  title         = {Concordance and Discordance in Cosmology},
  eprint        = {1806.04649},
  archivePrefix = {arXiv},
  primaryClass  = {astro-ph.CO},
  journal       = {Phys. Rev. D},
  volume        = {99},
  number        = {4},
  pages         = {043506},
  year          = {2019},
  doi           = {10.1103/PhysRevD.99.043506}
}

@article{RaveriZacharegkasHu:2019,
  author        = {Raveri, Marco and Zacharegkas, Georgios and Hu, Wayne},
  title         = {Quantifying concordance of correlated cosmological data sets},
  eprint        = {1912.04880},
  archivePrefix = {arXiv},
  primaryClass  = {astro-ph.CO},
  journal       = {Phys. Rev. D},
  volume        = {101},
  number        = {10},
  pages         = {103527},
  year          = {2020},
  doi           = {10.1103/PhysRevD.101.103527}
}

@article{Lemos:2020,
  author        = {Lemos, P. and Raveri, M. and Campos, A. and Park, Y. and Chang, C. and Weaverdyck, N. and others},
  title         = {Assessing tension metrics with Dark Energy Survey and Planck data},
  eprint        = {2012.09554},
  archivePrefix = {arXiv},
  primaryClass  = {astro-ph.CO},
  journal       = {Mon. Not. Roy. Astron. Soc.},
  volume        = {505},
  number        = {4},
  pages         = {6179--6194},
  year          = {2021},
  doi           = {10.1093/mnras/stab1670}
}

@article{Cowan:2010,
  author        = {Cowan, Glen and Cranmer, Kyle and Gross, Eilam and Vitells, Ofer},
  title         = {Asymptotic formulae for likelihood-based tests of new physics},
  eprint        = {1007.1727},
  archivePrefix = {arXiv},
  primaryClass  = {physics.data-an},
  journal       = {Eur. Phys. J. C},
  volume        = {71},
  pages         = {1554},
  year          = {2011},
  doi           = {10.1140/epjc/s10052-011-1554-0}
}
\end{document}